\documentclass[twocolumn,showpacs,superscriptaddress,amssymb,hyperlink]{revtex4-1}
\usepackage{amsmath,amssymb,wasysym,ytableau}
\usepackage{graphicx}
\usepackage{bm}
\usepackage{braket, slashed}
\usepackage{xcolor}
\usepackage{verbatim}
\usepackage{float}
\usepackage{caption,subcaption}
\usepackage[T1]{fontenc}
\usepackage{ulem}
\usepackage[colorlinks=true ,urlcolor=blue,citecolor=blue,urlbordercolor={0 1 1}]{hyperref}

\renewcommand{\vec}{\boldsymbol}
\newcommand{\tr}{\operatorname{Tr}}

\begin{document}
\title{Unnecessary quantum criticality in $SU(3)$ kagome magnets}
\author{Yunchao Zhang} 
\affiliation{Department of Physics, Massachusetts Institute of Technology, Cambridge MA 02139-4307,  USA}

\author{Xue-Yang Song}
\affiliation{Department of Physics, Hong Kong University of Science and Technology, Clear Water Bay, Hong Kong, China}

\author{T. Senthil}
\affiliation{Department of Physics, Massachusetts Institute of Technology, Cambridge MA 02139-4307,  USA}

\date{\today} 

\date{\today}   

\begin{abstract}
Algebraic/Dirac spin liquids (DSLs) are a class of critical quantum ground states that do not have a quasi-particle description.
DSLs and related spin liquid phases often arise in strongly frustrated quantum spin systems, in which strong correlations and quantum fluctuations among constituent spins persist down to zero temperature. 
In this work, we analyze Mott insulating phases of $SU(3)$ fermions on a kagome lattice which may realize a DSL phase, described at low energies by $(2 + 1)d$ quantum electrodynamics (QED$_3$) with $N_f=6$ Dirac fermions. 
By analyzing the action of physical symmetries on the operators of the QED$_3$ theory,  we conclude that the low energy DSL is a quantum critical point that can be accessed by tuning a single microscopic parameter. 
Aided by the emergent symmetry and anomalies of the low energy effective theory, we conjecture and present supporting arguments that the $SU(3)$ Kagome magnet DSL is an unnecessary quantum critical point, lying completely within a single phase.

\end{abstract}

\maketitle

\section{Introduction}
Quantum spin liquids are a class of quantum phases of matter that have appealing features from long-range quantum entanglement to the absence of Landau order \cite{anderson_resonating_1973,balents_spin_2010, knolle_field_2019,savary_quantum_2017,Zhou_2017,broholm_quantum_2020, hermele_algebraic_2005,ran_projected-wave-function_2007,hermele_properties_2008}.
These phases host a wide variety of correlated phenomena, including emergence of anyons, fractionalized excitations, and emergent gauge theory.

Gapped spin liquids, many of which possess intrinsic topological order, and certain classes of gapless spin liquids, such as the Kitaev spin liquid with non-interacting Majorana fermions \cite{Kitaev_2006}, are amenable to a quasiparticle formulation at low energy. 
However, there are types of quantum spin liquids that admit no quasi-particle description.
An example of such a quantum spin liquid is when the infra-red (IR) theory is a conformal field theory (CFT) fixed point at low energy.
Perhaps the simplest example is the spin-$1/2$ antiferromagnetic Heisenberg model in one dimension, which is described by the $SU(2)_1$ CFT \cite{witten_non-abelian_1984,affleck_exact_1986,affleck_critical_1988}.

In $(2+1)d$, the simplest known example is the $U(1)$ Dirac spin liquid (DSL) \cite{hermele_stability_2004,hermele_algebraic_2005,senthil_2005,ran_projected-wave-function_2007,hermele_properties_2008, senthil_competing_2006},
described by coupling $N_f$ gapless Dirac fermions $\psi$ to an emergent $U(1)$ gauge field $a$,
\begin{equation}
\label{eq:DSL}
    \mathcal{L}_{DSL}=\sum_{i=1}^{N_f} \overline{\psi}_ii\slashed{D}_a\psi_i+\frac{1}{4e^2}f_{\mu\nu}f^{\mu\nu},
\end{equation}
where $f_{\mu\nu}=\partial_{\mu}a_{\nu}-\partial_{\nu}a_{\mu}$. For sufficiently large $N_f$ ($ \geq 4$ according to numerical calculations), this theory flows in the IR to a CFT \cite{Appelquist_prl_1988}. The existence of an IR CFT for $N_f\ge 4$ is also consistent with bootstrap studies \cite{chesterjhep2016,li_2022_jhep,poland_prd_2022,he_scipost_2022,li2022plb}.
The most important local operators in this theory are built from monopole operators, which insert $q$ units of $2\pi$ $U(1)$ flux into the system at a spacetime point.
At the CFT fixed point, the scaling dimension of the fundamental monopoles ($q=1$) is given by \cite{dyer_monopole_2013}
\begin{equation}
    \Delta_{1}=0.265N_f-0.0383+\mathcal{O}(1/N_f),
\end{equation}
from a $1/N_f$ expansion.
For the DSL we examine in this work, $N_f=6$ and $\Delta_1\sim 1.55$.
Therefore, the monopole is a relevant operator and requires special analysis.
If the fundamental monopole is allowed by microscopic symmetries, the DSL will be unstable to monopole proliferation.
Note that at sufficiently large $N_f\gtrsim 12$, the monopoles will all be irrelevant \cite{borokhov2002topological, dyer_monopole_2013}. Thus the DSL will be a stable \cite{hermele_stability_2004} quantum spin liquid phase of matter for sufficiently large-$N_f$.

A large body of calculations and simulation suggest the DSL to be the ground state for a variety of lattice spin systems \cite{ran_projected-wave-function_2007,iqbal2013gapless,iqbal2016j1j2,he2017signatures,hu2019dsl}.
Furthermore, there are many material platforms that have shown evidence \cite{Wen2019experimental,ding_prb_2019,bordelon_nat_2019,zeng_prb_2022,zeng_nat_2024,ma_prb_2024,bag_prl_2024} for realizing a DSL state.
Fundamentally, these platforms are described as electronic Mott insulators where the quantum spin liquid is formed by interactions between localized spin moments in a lattice.
Attempts to move beyond this paradigm by generalizing the spin symmetry to $SU(N)$ in electronic spin systems have been challenging.
Such generalizations usually involve pairing spin and orbital degrees of freedom so locally, electrons can carry a representation of $SU(N)$ for $N>2$ even, but the resulting quantum spin-orbital liquids are often not fully $SU(N)$ symmetric.
Moreover, spin-orbit coupling can break the individual spin and orbit-space symmetries, although there are special cases in which \textit{strong} spin-orbit coupling actually leads to enhanced symmetry \cite{yamada2018emergent}.
However, as noted earlier, the stability of the DSL state and resulting low energy CFT is greatly enhanced for larger $N$ (in the systems relevant to our study, $N_f\propto N$), motivating the study of $SU(N)$-symmetric spin liquids for larger $N$.

Concretely, an $SU(N)$ spin liquid can potentially be realized in ultracold atomic systems by manipulating nuclear spins.
Through loading $N$-color atoms on optical lattices, one can can simulate $SU(N)$-symmetric interactions in a defect-free and fully controllable environment, providing an experimental platform to engineer and probe exotic phases not yet discovered (or maybe even accessible) in real materials hosting electronic systems. More generally, ultracold gases of alkaline-earth-like atoms prepared in the lowest two electronic states can obey $SU(2I+1)$ symmetry, where $I$ is the nuclear spin \cite{wu_prl_2003,hermele2009mott,gorshkov2010two,hermele2011sun,Cazalilla_reports_2014,Scazza_nature_2014}.
With atoms such as $^{87}$Sr, the nuclear spin is as large as $9/2$, leading to an $SU(10)$ symmetric system.
Furthermore, by choosing an appropriate initial state and using the fact that the total number of atoms with a given nuclear spin $m$ is conserved, atoms with large $I$ such as $^{87}$Sr can realize the interactions of atoms with lower $I$, such that an $SU(N)$ symmetric system for any $N\le 2I+1$ can be simulated \cite{gorshkov2010two}.

In this work, we will focus on Mott insulators of an $SU(3)$ fermion system with one atom per site on the kagome lattice, which can be programmed on a cold atoms platform \cite{santos_prl_2004,jo_PRL_2012}.
To realize $SU(3)$ symmetric interactions, one can use alkaline-earth-like atoms with nuclear spin $I\ge 1$, such as $^{173}$Yb$(I=5/2)$.
We will study a Dirac spin liquid that can potentially occur in such a Mott insulator, and ask where it is situated in the phase diagram. 

Through analyzing the quantum numbers of relevant monopole operators, we show there is a relevant monopole operator that transforms trivially under all microscopic symmetries.
Therefore, this DSL is unstable, as the presence  of this monopole will lead to an instability, possibly towards various symmetry breaking orders. However we argue that there is only a single such relevant operator at the QED$_3$ fixed point that is allowed by the microscopic symmetries. Thus this fixed point will appear as a quantum critical point in the phase diagram of the $SU(3)$ Kagome magnet. 

We further argue that the DSL on the kagome lattice is likely an ``unnecessary" quantum critical point (QCP) \cite{bi_prx_2019}.
Unnecessary QCPs describe transitions within  a single phase of matter, in contrast to a conventional QCP that describes the phase transition between two \textit{different} phases upon the tuning of a relevant perturbation. For other examples, see Refs.~\cite{bi_prx_2019,xu_prb_2020,parameswaran_prl_2023,senthil2024deconfined,zhang2024diracspinliquidunnecessary,prakash2024charge}.
In our previous work, we showed that the Dirac spin liquid also plausibly describes an unncessary QCP within the Neel (or valence bond solid) phase of spin-$1/2$ square lattice magnets \cite{zhang2024diracspinliquidunnecessary}. 
Thus the present work adds the $SU(3)$ Kagome lattice DSL as another example of such unnecessary QCPs.

Our arguments will rely on an analysis of the relation between the microscopic symmetries of the lattice model and the emergent symmetries of the IR continuum field theory. We will analyze the 't Hooft anomalies of the latter and the constraints they impose on the low energy physics of the perturbed QED$_3$ theory. We will find that the continuum field theory defined by the perturbing the DSL CFT with its single relevant symmetry-allowed operator has an anomaly that precludes it from flowing to a trivial gapped phase. Further, the anomaly even precludes any gapped symmetric phase even allowing for the possibility of topological order, a phenomenon known as symmetry-enforced gaplessness \cite{wang_prb_2014}. We conjecture that the  endpoint of the renormalization group (RG) flow away from the DSL is a symmetry broken phase. We show that the same such phase will be obtained for either sign of the relevant perturbation, which then  establishes  that this DSL is an unnecessary critical point. 

Our works adds to the growing number of examples of unnecessary quantum critical points which may thus be not an uncommon phenomenon in quantum many body physics. 

The rest of this paper is organized as follows.
In Sec.~\ref{sec:mft} we introduce the kagome lattice mean field ansatz and discuss aspects of the $U(1)$ DSL state, focusing on the low energy effective field theory and the pivotal role of monopole operators in the critical theory.
In Sec.~\ref{sec:sym}, we will explain the connection between the lattice symmetries and the emergent symmetries of the infrared continuum QED$_3$ theory.
Analyzing the embedding of the lattice symmetries into the continuum to constrain transformation properties of the monopole operators, we will conclude the DSL is unstable due to a monopole trivial under all symmetries.
The resulting DSL is then not a stable phase, but a quantum critical point.
In Sec.~\ref{sec:anomalies} we calculate the quantum anomalies of the QED$_3$ QCP, both in the absence and presence of monopole proliferation.
Through the nature of the anomaly, we argue that the DSL is a unnecessary quantum critical point, tuned by the proliferation of the
singlet monopole.
This was previously found in the $N_f=4$ DSL on the square lattice \cite{zhang2024diracspinliquidunnecessary}, so we find that unnecessary criticality may be a feature common to critical points described by QED$_3$.
We conclude in Sec.~\ref{sec:conclusion} with some discussion and further outlook.

\section{Candidate DSL on on the kagome lattice}
\label{sec:mft}
As discussed in the previous section, by externally trapping ultracold alkaline-earth atoms, one can realize $SU(3)$-symmetric interactions on the kagome lattice.
A model Hamiltonian that describes such systems is given by the $3$-flavor Hubbard model \cite{gorshkov2010two,Cazalilla_reports_2014}, 
\begin{equation}
\label{eq:hubbard}
    H=-\sum_{\braket{\vec{r},\vec{r}'}}\sum_{\alpha=1}^3 (t_{\vec{r}\vec{r}'}c^{\dagger}_{\vec{r}\alpha}c_{\vec{r}'\alpha}+h.c.)+U\sum_{\vec{r}}\left(\sum_{\alpha=1}^3n_{\vec{r}\alpha}\right)^2.
\end{equation}
where $U>0$ captures a finite repulsive on-site density-density interaction and $\alpha$ labels the $SU(3)$ spin index.
The kagome lattice has a three-site unit cell, and we adopt a convention in which the primitive lattice vectors are defined as
\begin{equation}
    \vec{R}_1=2a\hat{x},\quad
    \vec{R}_{2,3}=2a\left(\frac{\pm\hat{x}+\sqrt{3}\hat{y}}{2}\right),
\end{equation}
leading to a hexagonal reciprocal lattice with corners at $\vec{K}=\left(\frac{2\pi}{3a},0\right)$ and $\vec{K}'=-\vec{K}$.

Focusing on $1/3$ filling, what are the expected phases as we tune $U$?
In the
the free limit $U\rightarrow0$, one obtains a simple tight binding model.
Taking a uniform hopping amplitude $t_{\vec{r}\vec{r}'}=t$,
\begin{equation}
\label{eq:hopping}
    H=-t\sum_{\braket{\vec{r},\vec{r}'}}\sum_{\alpha=1}^3 (c^{\dagger}_{\vec{r}\alpha}c_{\vec{r}'\alpha}+h.c.),
\end{equation}
we find a flat upper band and two lower bands that intersect at Dirac points at $\vec{K}$ and $\vec{K}'$.
At $1/3$ filling, the Fermi level then passes right through the Dirac points, yielding a Dirac semimetal state.
The tight binding spectrum $\epsilon(\vec{k})$ of Eq.~\eqref{eq:hopping} is outlined in Appendix~\ref{app:mft} and shown in Figure~\ref{fig:dispersion}.
\begin{figure}[b]
\centering
\subcaptionbox{}
{\includegraphics[width=\columnwidth]{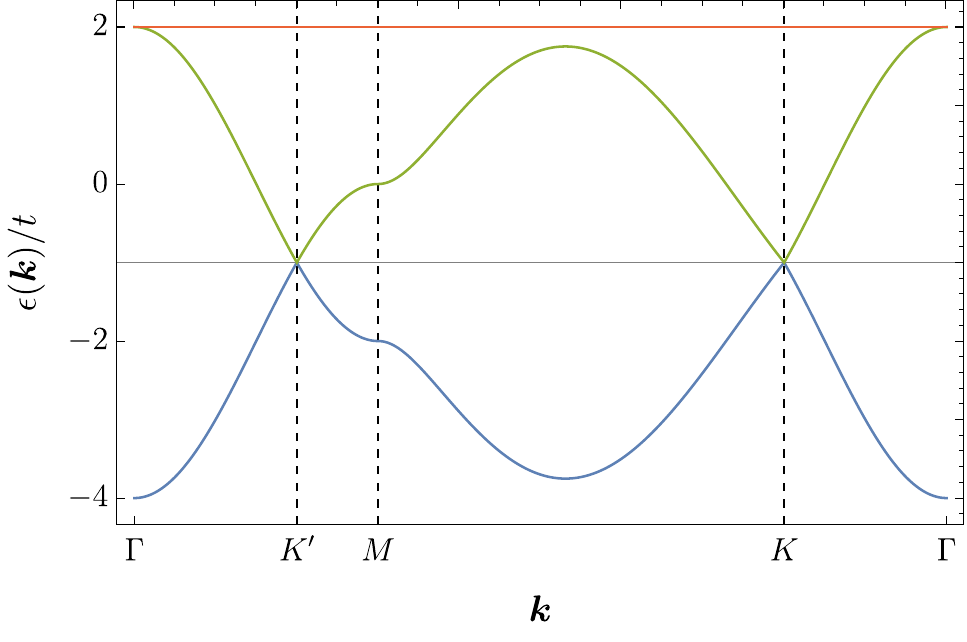}}
\subcaptionbox{}
{\includegraphics[width=0.8\columnwidth]{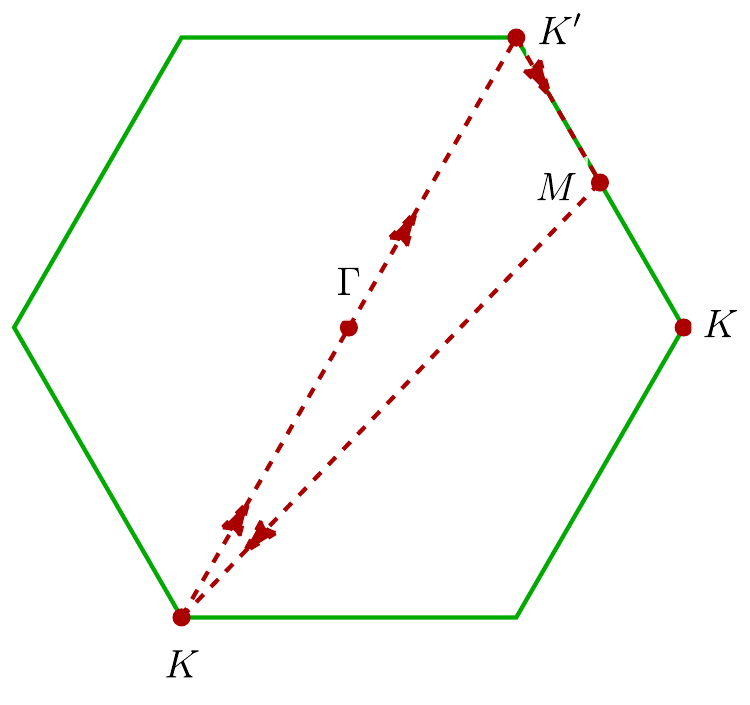}}
\captionsetup{justification=raggedright}
\caption{(a) Dispersion of the tight binding spinon Hamiltonian, Eq.~\eqref{eq:hopping}, with $t=1$. The Brioullin zone path is shown in (b).
We observe an upper flat band and lower bands which intersect at Dirac points at $\vec{K}$ and $\vec{K}'$.}\label{fig:dispersion}
\end{figure}
As short range interactions are irrelevant at the Dirac semimetal fixed point, we see the Dirac semimetal is a stable phase for weak Hubbard coupling $U$.

For sufficiently large $U$, the resulting state is a Mott insulator.
The large $U$ limit is described by an $SU(3)$ antiferromagnetic Heisenberg Hamiltonian,
\begin{equation}
    H_{eff}\sim\frac{t^2}{U}\sum_{\langle\vec{r},\vec{r}'\rangle}P_{\vec{r},\vec{r}'},
\end{equation}
where $P_{\vec{r},\vec{r}'}$ permutes the fermions between two nearest neighbor sites $\vec{r}$ and $\vec{r}'$.
Each site hosts an $SU(3)$ spin transforming in the fundamental representation $\mathbf{3}$.

\begin{figure}[t]
\captionsetup{justification=raggedright}
{\includegraphics[width=0.9\columnwidth]{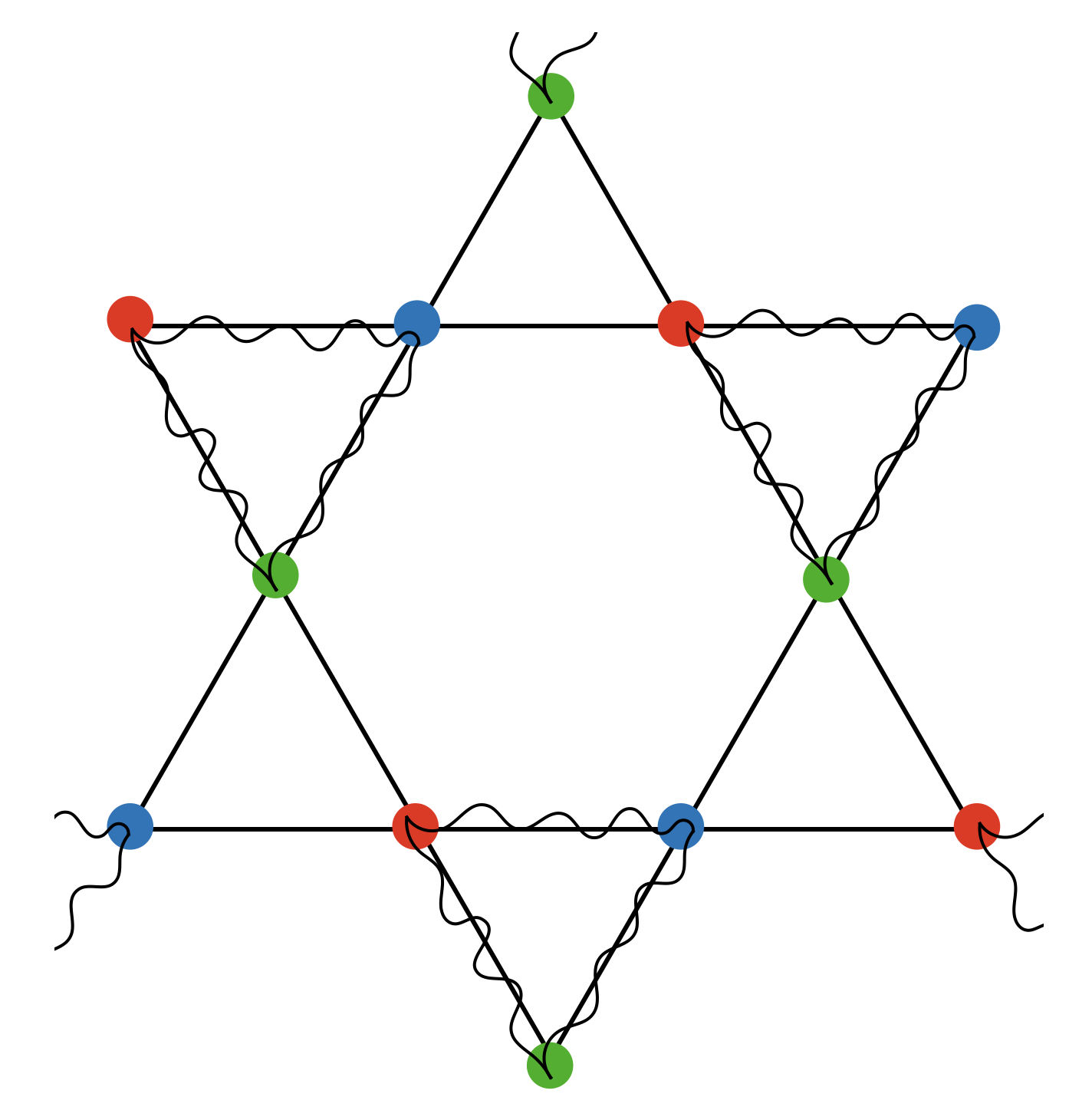}}
\caption{
A schematic illustration of the ground state in the $U\rightarrow \infty$ limit, known as a simplex VBS state, with each collection of $SU(3)$ spins forming a trimer singlet that covers the entire kagome lattice.
The ground state preserves $C_3$ and lattice translation symmetries, with $C_6$ (and consequently, lattice inversion) spontaneously broken.}
\label{fig:vbs}
\end{figure}

Numerical studies \cite{corboz_prb_2012,liu_prb_2015,changlani_prb_2015,xu_prb-2023} support that the ground state will be be a
valence bond crystal (VBS)-like state that preserves $C_3$ and lattice translations, as a single unit cell can realize an $SU(3)$ singlet state, $\otimes^3\mathbf{3}=\mathbf{1}\oplus\mathbf{8}^2\oplus\mathbf{10}$.
This ground state of trimers, shown in Figure~\ref{fig:vbs}, is a specific case of the more general $N$-simplex solid state for $SU(N)$ antiferromagnets, in which $N$-site $SU(N)$ singlets cover the plaquettes of an $N$-partite lattice \cite{arovas_prb_2008,hermele2011sun}, generalizing the valence bond solid state as
that singlets do not populate two-site bonds, but instead $N$-site clusters.

As such a trimerized state on the kagome lattice spontaneously breaks inversion symmetry, the resulting ground state manifold is doubly degenerate.
At intermediate $U/t\sim 1$, the perturbatively obtained antiferromagnetic Hamiltonian is no longer a suitable description of the system as higher order ring-exchange terms become important \cite{macdonald_prb_1988}.
The ground state is not well understood. One possibility is that there is a direct Gross-Neveau transition between the Dirac semi-metal and the VBS state driven by opening a fermion bilinear mass gap. Here however, we will explore an alternate possibility in which   near the Mott transition, a Dirac spin liquid may appear, and instead of electrons, fractionalized $SU(3)$ fermionic ``spinons'' hop on the kagome lattice with zero background flux.
This is the state we will study, and to do so, we employ the parton decomposition of $SU(3)$ spins $\vec{S}$
into fermionic spinons $\vec{f}$ by writing
\begin{equation}
\label{eq:spinon}
\vec{S}_r^{A}=\sum_{\alpha,\eta=1}^3f_{r,\alpha}^{\dagger}T_{\alpha\eta}^Af_{r,\eta},
\end{equation}
where $r$ is the site index.
The $f_{r,\alpha}^{(\dagger)}$ are second quantized spinon field operators at site $r$ with $SU(3)$ flavor $\alpha\in\{1,2,3\}$.
Note that the $f$ operators transform as $SU(3)$ fundamentals, while the $T^A$ are the matrix representation of the Hermitian $SU(3)$ generators acting on the fundamental representation, normalized by the condition $\operatorname{Tr}[T^AT^B]=\frac{1}{2}\delta^{AB}$. $A$ is an adjoint index, which ranges in $\{1,2,\dots,8\}$.

In order to reproduce the physical Hilbert space, the relation in Eq.~\eqref{eq:spinon}
comes with an additional constraint that $\sum_{\alpha}f_{r,\alpha}^{\dagger}f_{r,\alpha}=1$.
We then perform a mean field decomposition of the Hubbard Hamiltonian, keeping all spinon hopping terms while preserving the constraint only at the expectation value level.

The result is a mean field Hamiltonian of fermionic particles on a kagome lattice at $1/3$ filling,
\begin{equation}
\label{eq:hopping_spinon}
    H=-t\sum_{\braket{\vec{r},\vec{r}'}}\sum_{\alpha=1}^3 (f^{\dagger}_{\vec{r}\alpha}f_{\vec{r}'\alpha}+h.c.),
\end{equation}
or one atom per lattice site.
Note in the above, the fermionic $f$ operators represent the fractionalized $SU(3)$ spinons, not the physical electrons.
Each site hosts three degenerate atomic states (which can be interpreted as spin states, orbitals, etc.) originating from the three fermion flavors. The dispersion is the same as in Eq.~\eqref{eq:hopping} and Figure~\ref{fig:dispersion}.

We observe that the decomposition in Eq.~\eqref{eq:spinon} as realized in 
Eq.~\eqref{eq:hopping_spinon} has a $U(1)$ gauge redundancy generated by the constraint $\sum_{\alpha}f_{r,\alpha}^{\dagger}f_{r,\alpha}=1$ that maps each $f_R$ by a site-dependent phase factor, $f_r\rightarrow e^{i\varphi_r}f_r$.
In order to restore gauge invariance, we will introduce an emergent $U(1)$ gauge field $a_{\mu}$ that will appear in the low energy field theory \footnote{Shortly, we will find it convenient to place the IR continuum field theory on an arbitrary oriented space-time manifold. We will then regard $a$ as a spin$_c$ connection rather than an ordinary $U(1)$ gauge field.}.

\subsection{Low energy field theory}
Performing an expansion of the momentum-space Hamiltonian around the Dirac points, $H(\pm\vec{K}+\vec{q})-H(\pm\vec{K})$, and with a suitable redefinition of the fermion fields, we obtain an effective low energy Dirac Hamiltonian
\begin{equation}
\mathcal{H}_{Dirac}=v_F\int \frac{d^2\vec{q}}{(2\pi)^2}
\psi_{\alpha a}^{\dagger}(q_x\tau^1+q_y\tau^y)\psi_{\alpha a}.
\end{equation}
The first index labels the $SU(3)$ spin, the second index labels the Dirac node ($\pm\vec{K}$), and we have made the spinor index implicit.

Focusing on the $6$ Dirac nodes near the Fermi energy coming from the $3$ spin degrees of freedom and $2$ valleys, one can describe the low energy excitations with the effective Lagrangian,
\begin{equation}
\label{eq:qed_action}   \mathcal{L}=\mathcal{L}_{QED_{3}}+\Delta\mathcal{L}=\sum_{i=1}^6 \overline{\psi}_i\slashed{D}_a\psi_i-\frac{1}{4e^2}f_{\mu\nu}^2+\Delta\mathcal{L},
\end{equation}
where $\psi$ are two component Dirac fermions with flavor index $i\in\{1,2,\dots,6\}$ and $(\gamma_0,\gamma_1,\gamma_2)=(-i\tau^3,\tau^2,-\tau^1)$ are the gamma matrices in $(2+1)d$.
The adjoint Dirac fermion is $\overline{\psi}^{\alpha}=(\psi^{\dagger})^{\alpha}(-i\gamma^0)$.
Lastly, we have defined $f_{\mu\nu}=\partial_{\mu}a_{\nu}-\partial_{\nu}a_{\mu}$ to be the curvature of the emergent $U(1)$ gauge field $a$.

$\mathcal{L}_{QED_{3}}$ is exactly the Lagrangian for $(2+1)d$ quantum electrodynamics (QED$_3$) with $N_f=6$ Dirac fermions, which includes minimal coupling of the emergent gauge field to the spinons, in addition to a Maxwell term for the gauge field.
$\Delta\mathcal{L}$ includes additional operators in QED$_3$ that are allowed by the microscopic symmetries of lattice Hamiltonian.

\subsection{Monopole operators}
As mentioned in the introduction, in the absence of $\Delta\mathcal{L}$, Eq.~\eqref{eq:DSL}
flows to an interacting CFT fixed point \cite{appelquist_1985,appelquist_1986,Appelquist_prl_1988,hermele_stability_2004,karthik_scale_2016,karthik_no_2016,dyer_monopole_2013}.
However, the most relevant operators in the renormalization group sense are the lowest charge monopole operators with scaling dimension \cite{borokhov2002topological,dyer_monopole_2013} $\Delta_1\sim 1.55+\mathcal{O}(\frac{1}{N_f})$ for $N_f=6$.
The presence of these monopoles is therefore the most important factor affecting the stability and proximate phases of the DSL.
To determine if monopoles are allowed in the action, we must determine their quantum numbers, which are constrained by the symmetries of both the infrared (IR) continuum QED$_3$ theory and the ultraviolet (UV) lattice ansatz.
Before delving into the action of symmetry on the monopoles, we first review some background on monopole operators in QED.

To begin, the QED$_3$ action in Eq.~\eqref{eq:DSL} generally allows for topologically nontrivial configurations of the gauge field.
These can be described by local monopole operators $\mathcal{M}_q$, which insert an integer $q$ units of $2\pi$ $U(1)$ flux into the system at a spacetime point \cite{borokhov2002topological}.
Identically, monopole operators $\mathcal{M}^{\dagger}_q$ can be viewed as non-trivial topological configurations of the gauge field $a$, which carry charge $q$ under a global $U(1)$ symmetry with conserved current equal to the magnetic flux
\begin{equation}
    j^{\mu}=\frac{1}{2\pi}\epsilon^{\mu\nu\lambda}\partial_{\nu}a_{\lambda}.
\end{equation}
We denote this flux symmetry by $U(1)_{top}$.
In the absence of the Dirac fermions and other matter fields, the monopole perturbation will gap out the photon field and confine gauge charges \cite{polyakov1977quark}.
In the presence of Dirac fermions, the bare monopole operators $\mathcal{M}^{\dagger}_q$ must be dressed by fermion zero modes in order to be gauge invariant.
The monopoles can be viewed as the basic building blocks of QED$_3$, in that
other local operators, including fermion billinears $\overline{\psi}_i\psi_j$, can identified with composites of the gauge-invariant monopole operators.

As we are interested in fluctuations around the zero-flux vacuum, we will focus on the fundamental ($q=1$) monopoles with scaling dimension \cite{borokhov2002topological,dyer_monopole_2013}
\begin{equation}
    \Delta_{1}=0.265N_f-0.0383+\mathcal{O}(1/N_f)\sim 1.55. 
\end{equation}
Monopoles with higher charge have higher scaling dimension and are irrelevant at the QED$_3$ CFT.
For our intuitive understanding, it is useful to transition into the large $N_f$ limit, in which fluctuations of the gauge field are suppressed and the bare monopole operator $\mathcal{M}^{\dagger}_1$ creates a uniform $2\pi$ flux.
The properties of monopole operators can then be deduced from a theory of free Dirac fermions in a constant background magnetic flux.
By the Atiyah-Singer index theorem, each Dirac cone contributes a single zero-energy mode.
Gauge invariance constrains that exactly half of the zero modes must be filled.
In the case of $N_f=6$ as in Eq.~\eqref{eq:qed_action}, we obtain a total of 
$\binom{6}{3}=20$ monopoles, each of which has three zero modes occupied. 
We can represent the gauge invariant monopole operators schematically by 
\begin{equation} \epsilon^{ijk}f_{i}^{\dagger}f_{j}^{\dagger}f_{k}^{\dagger}\mathcal{M}_1^{\dagger}.
\end{equation}

We have denoted $f^\dagger_i$ as the creation operator for the zero mode corresponding to the parent Dirac fermion $\psi_i$ and $\mathcal{M}_1^{\dagger}$ as the bare monopole operator which simply inserts the unit flux without filling any zero mode.
Analogously, antimonopole operators, which insert unit $-2\pi$ flux, can be defined as the Hermitian conjugates of the monopole operators.
The zero modes transform under the flavour symmetry that mixes the $6$ Dirac fermions, but as they have vanishing angular momentum in the flux background, the zero modes are singlets under all Lorentz transformations.
We also remark that in this large $N_f$ limit, the monopoles are irrelevant.

Note that other operators are potentially allowed in the $N_f=6$ theory as long as they transform trivially under the UV symmetries.
However, these operators are most likely irrelevant, as shown in $1/N_f$ expansions and numerical simulation.
The most relevant, both physically and in the RG sense, include couplings of monopoles to fermion billinears and strength two monopoles.
The first type of operator is equivalent to exciting a fundamental monopole.
At leading order, it involves transitioning a Dirac zero mode $f^{\dagger}$ from the $n=-1$ Landau level to $n=1$, and the resulting operator has an irrelevant scaling dimension $\Delta_{1}+2\sqrt{2}\sim 4.38>3$ from large $N_f$.

The strength two monopole can be created by the double insertion of a strength one monopole or, analogous to the case of a $2\pi$ flux, by dressing a bare $4\pi$ flux with $6$ out of the $12$ Dirac fermion zero modes. In either case, the strength two monopole scaling dimension estimated from large $N_f$ is $\Delta_2=0.673N_f-0.194\sim 3.84$.
While the closeness of $\Delta_2$ to $3$ suggests that it could be marginal or relevant due to higher order $1/N_f$ corrections, recent Monte Carlo calculations \cite{karthik_prd_2024} estimate $\Delta_2=3.73(34)$ for $N_f=4$.
As these monopoles become more irrelevant with increasing $N_f$, these numerics strongly suggest that the strength two monopole in $N_f=6$ is irrelevant as well. 

\section{Symmetries of the DSL}
\label{sec:sym}
In this section, we will illustrate how symmetries of the microscopic theory constrain the allowed monopole operators, following the methods employed in Refs.~\cite{Song_2019,Song_2020}.
While we are only considering a specific lattice ansatz, the methods used illustrate a more general framework.
Whenever a lattice model produces as emergent critical field theory in the IR limit, the operators in the IR field theory should be interpreted physically as coarse-grained, dressed lattice operators.
The correspondence is made by matching the symmetries of lattice operators with those of the IR field theory.
At a more abstract level, there is some embedding/group homomorphism ($\varphi$) of the lattice symmetries $G_{UV}$ into the symmetry group $G_{IR}$ of the IR theory,
\begin{equation}
    \varphi\;:\;G_{UV}\rightarrow G_{IR}.
\end{equation}
Our concern will be to find the symmetry action of $\varphi(G_{UV})$ on the fundamental monopole operators to find which are allowed from the UV lattice theory.

\subsection{Symmetries of the microscopic system: the group $G_{UV}$}
From the UV lattice model (which we take to be a spin model with spins transforming in the fundamental representation of $SU(3)$), the total microscopic symmetry is 
\begin{equation}
    G_{UV}= {PSU(3)_s} \rtimes\mathcal{T}\times G_{lattice},
\end{equation}
where $PSU(N)=SU(N)/Z(SU(N))$ is the projective special unitary group.
The point group symmetries of the kagome lattice, $G_{lattice}$, are generated by translations ($T_{1,2}$) along the primitive Bravais lattice vectors $\vec{R}_{1,2}$, rotations ($C_6$) by $2\pi/6$ around the origin of a hexagonal plaquette, and a reflection symmetry ($\mathcal{R}_{y}$) taking $y\rightarrow -y$.
Furthermore, we have $PSU(3)_s$ spin rotations and (antiunitary) time reversal $\mathcal{T}$.

Note that as there are 3 sites per kagome unit cell, and we have $SU(3)$ spins which can then form a singlet within each unit cell,  there is no Lieb-Schultz-Mattis (LSM) constraint coming from just the interplay between lattice translations and the $PSU(3)$ symmetry. Indeed, the construction of the translation-invariant VBS Mott insulator described earlier illustrates the absence of such a non-trivial LSM constraint. 

\subsection{Symmetries of  QED$_3$: the group $G_{IR}$}
We now review the symmetries of the continuum QED$_3$ theory which form $G_{IR}$.
The theory $\mathcal{L}_{QED_3}$ from Eq.~\eqref{eq:DSL} has a continuous Lorentz group symmetry \footnote{At the fixed point to which the QED$_3$ flows, we will also have conformal symmetry.}, $SO(2,1)_L$, 
and the familiar discrete Lorentz symmetry actions of time reversal $\mathcal{T}_{IR}$, partial reflection $\mathcal{R}_{IR}$, and charge conjugation $\mathcal{C}_{IR}$.
The \textit{IR} subscript is a reminder that these bare actions in $G_{IR}$ are not the same as the corresponding physical symmetry in $G_{UV}$.
The discrete symmetries can be chosen to act on the Dirac fermions as
\begin{alignat}{1}
\mathcal{T}_{IR}\;&:\; \psi(t,\vec{r})\rightarrow \gamma^1\psi^{\alpha}(-t,\vec{r}),\quad i\rightarrow -i,\label{eq:bareaction_dirac}\\
    \mathcal{R}_{IR}\;&:\;\psi(t,\vec{r})\rightarrow -i\gamma^2 \psi^{\alpha}(t,R\vec{r}),\\
    \mathcal{C}_{IR}\;&:\; \psi(t,\vec{r})\rightarrow (\overline{\psi}^{\alpha}(t,\vec{r})(i\gamma^1))^T.
    \label{eq:bareaction_dirac_end}
\end{alignat}
Note however that the Dirac fermions are not local operators; we will use them as a device to deduce the symmetry action 
Internally on the Dirac fermions, we also have (what is roughly) a global $SU(6)_f$ flavor symmetry,
$\psi_{\alpha}\rightarrow U_{\alpha\beta}\psi_{\beta}$ for $U\in SU(6)_f$.
However we should quotient out by the center of $SU(6)$ (which can be combined with a gauge transformation)  so that the theory naively has an internal symmetry group of $PSU(6)=SU(6)_f/\mathbb{Z}_6$. In addition we have a $U(1)_{top}$ symmetry associated with conservation of the gauge flux $da/2\pi$. This symmetry acts trivially on the Dirac fermions. 
The true internal symmetry group, as we will show below using the monopole operators, is not actually the simple product $PSU(6)\times U(1)_{top}$, but a quotient group.

As noted earlier, in the presence of a charge one monopole $\mathcal{M}_1^{\dagger}$, each Dirac mode contributes a single fermion zero mode $f_i$, 
which behaves as a Lorentz scalar while transforming as $\psi_i$ under $SU(6)_f$.
We will denote $\phi_A$ to be the monopole represented by the ``wave function''
\begin{alignat}{1}
    \phi_A^{\dagger}&=\mathcal{F}_{A}^{\dagger}\mathcal{M}^{\dagger},\\\nonumber
    \mathcal{F}_A^{\dagger}&=f_{[A_1}^{\dagger}f_{A_2}^{\dagger}f_{A_3]}^{\dagger},
\end{alignat}
where for convenience of notation, $A=[A_1,A_2,A_3]$ denotes an antisymmetrized multi-index defined by 
\begin{equation} 
T_{[a_{1}\dots a_{p}]}={\frac {1}{p!}}\epsilon_{a_{1}\dots a_{p}}\epsilon^{b_{1}\dots b_{p}}T_{b_{1}\dots b_{p}}.
\end{equation}

As the $f$'s transform in the fundamental of $SU(6)_f$ and $\mathcal{M}$ is an $SU(6)_f$ singlet, $\phi^{\dagger}$ will transform in the $3$-fold antisymmetric tensor product of the $SU(6)$ fundamental representation, which is the irreducible representation of $SU(6)$ whose Young tableau has one column and three rows.
This representation is self conjugate and symplectic, with an invariant bilinear given by 
\begin{equation}
    E_{AB}=E_{[A_1,\dots,A_3][B_1,\dots,B_3]}=\frac{1}{6}\epsilon_{A_1\dots A_3B_1\dots B_3}=E^{AB}.
\end{equation}
In the case $N_f\equiv 0\pmod{4}$ as was considered in previous works, the representation is instead orthogonal, and $E_{AB}$ is symmetric instead of antisymmetric  \cite{calvera2021theory}.
In either case, $E_{AB}$ should be thought of as acting on the space of monopole wave functions $\phi_A^{\dagger}$, equivalent to the $N_f/2$-fold antisymmetric product on $\mathbb{C}^{N_f}$.
In summary, we see that the $\phi^{\dagger}$ have charge one under the magnetic $U(1)_{top}$ symmetry, transform in the self-conjugate antisymmetric representation of $SU(6)_f$, and are Lorentz scalars.

The center of $SU(6)_f$ is made of matrices of the form $\omega^n\mathbb{I}_{6\times 6}$ for $\omega=e^{2\pi i/ 6}$, so the true faithful internal symmetry group acting on the monopoles is
\begin{equation}
    G_{IR}=\frac{SU(6)_f\times U(1)_{top}}{\mathbb{Z}_6},
\end{equation}
where $\mathbb{Z}_6$ is generated by the element $(\omega\mathbb{I}_{6\times 6},-1)\in SU(6)_f\times U(1)_{top}$.
The quotient action arises because $\omega\mathbb{I}\in SU(6)_f$ and a $\pi$ rotation in $U(1)_{top}$ are indistinguishable when acting on gauge invariant operators.
The full IR symmetry is then $G_{IR}$ in addition to the Lorentz and discrete symmetries such as time reversal, reflection, and charge conjugation.

To define the action of the discrete, infrared Lorentz symmetries on the monopole operators, we must determine how they act on the $U(1)_{top}$ and $SU(6)_f$ symmetry.
First, we observe that for all three discrete actions, $U(1)_{top}$ charge is reversed, so $\phi$ is mapped to its Hermitian conjugate.
$\mathcal{T}_{IR}$ and $\mathcal{R}_{IR}$ do not interact with the $SU(6)_f$ representation of the monopoles, but $\mathcal{C}_{IR}$ changes an $SU(6)_f$ representation to its complex conjugate.
Choosing appropriate phases allows us to define
\begin{alignat}{1}
\label{eq:bareaction_monopole}
    \mathcal{T}_{IR}\;&:\; \phi_A^{\dagger}\rightarrow iE_{AB} \phi^B,\\\nonumber
    \mathcal{R}_{IR}\;&:\; \phi_A^{\dagger}\rightarrow -E_{AB} \phi^B,\\\nonumber
    \mathcal{C}_{IR}\;&:\; \phi_A^{\dagger}\rightarrow i \phi^B,
\end{alignat}
where phases up to a $U(1)_{top}$ action were chosen so $\mathcal{T}_{IR}^2=(\mathcal{C}_{IR}\mathcal{R}_{IR})^2=1$ on the monopole operators \cite{calvera2021theory}.
These phase ambiguities are a matter of convention. 

\subsection{Embedding $G_{UV}$ into $G_{IR}$}
The parton construction leading to the QED$_3$ theory enables us to deduce how $G_{UV}$ embeds into $G_{IR}$. We begin by noting that 
as the fermionic spinons are not gauge invariant operators, the microscopic symmetries need not act linearly on them.
This is formalized with the notion of the spin liquid's projective symmetry group \cite{Wen2002quantum}.
The microscopic symmetries can then be written in terms of elements of the projective symmetry group that transform the physical spin operators $\vec{S}$ as desired.
However, the projective symmetry group also has a subgroup, termed the invariant gauge group (IGG), that leaves the physical spin operators invariant.
Consequently, symmetry actions on the spinons are only defined modulo elements of the IGG.
The IGG of our mean field ansatz is exactly the $U(1)$ gauge group emergent from the parton decomposition.
Once the microscopic symmetry actions on the partons are defined, they can be projected down to the IR degrees of freedom, yielding $\varphi(G_{UV})$.

\begin{figure}[t]
\captionsetup{justification=raggedright}
{\includegraphics[width=\columnwidth]{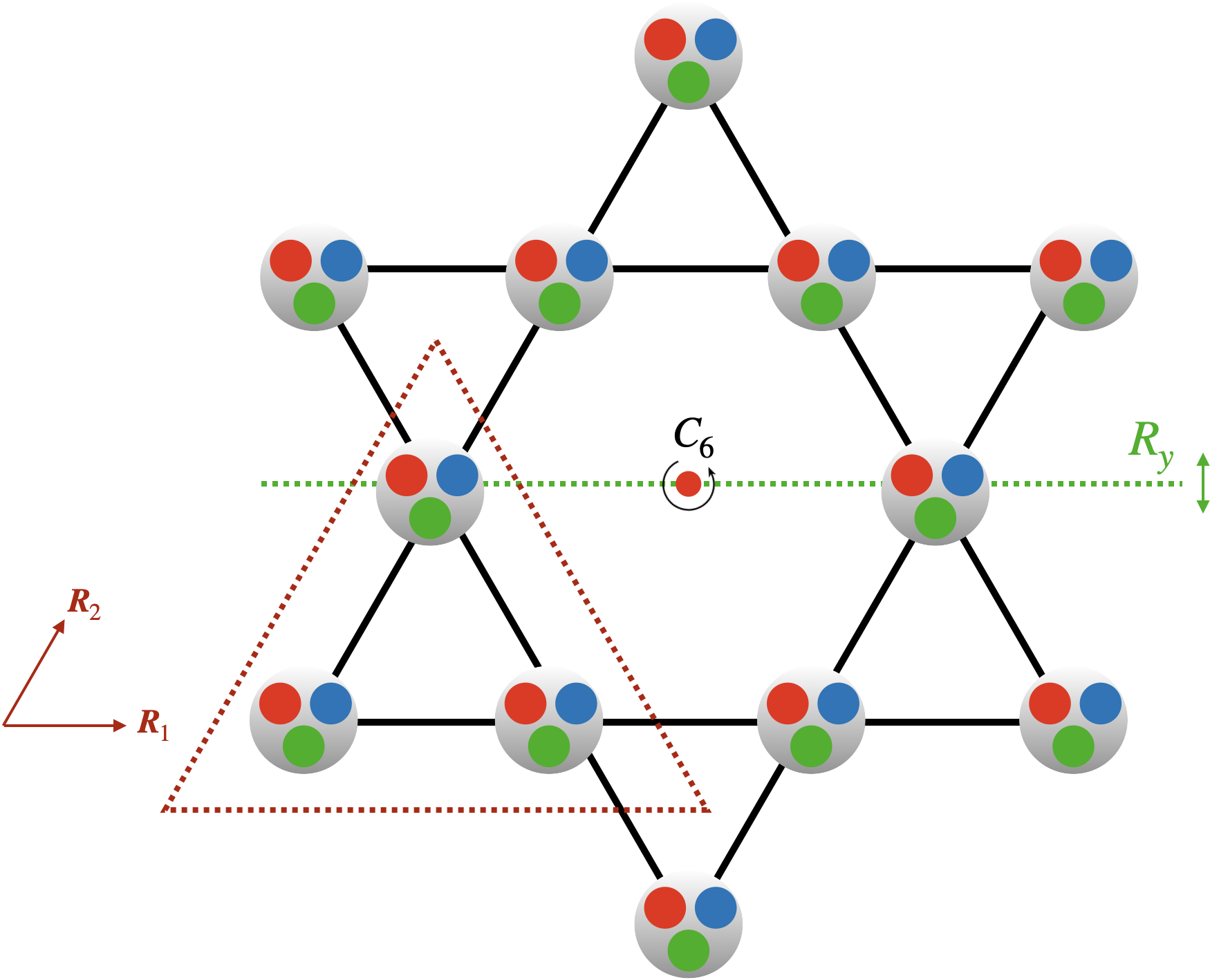}}

\caption{
Conventions for the kagome lattice symmetries.
The choice of unit cell is outlined in red, and the lattice space group generators are $C_6$, $\mathcal{R}_y$, and translations along the primitive lattice vectors $\vec{R}_{1,2}$.}

\label{fig:sym}
\end{figure}

Due to the simplicity of our spin liquid ansatz, the kagome lattice symmetries can be realized linearly on the spinons.
The induced symmetry action on the Dirac fermions $\psi$ is listed in Table~\ref{table:symmetries} and shown in Figure~\ref{fig:sym}.
We have defined $\mu$ to act on the Dirac node index, $\tau$ to act on the Lorentz (spinor) index, and time reversal to include complex conjugation.
We see these symmetries forbid all mass terms $\psi^{\dagger}\mu^i\tau^j\psi$.

\begin{table}[h]
\begin{tabular}{|l|l|}
\hline
Symmetry                            & Representation \\ \hline
$T_{1,2}$                               & $\psi\rightarrow e^{i\vec{K}\cdot\vec{R}_{1,2}\mu^3 }\psi$  \\
$C_6$               & $\psi\rightarrow i\mu^1e^{-\frac{i\pi}{6}\tau^3}\psi$           \\
$\mathcal{R}_{y}$                   & $\psi\rightarrow \mu^3\tau^1\psi$           \\
$\mathcal{T}$                       & $\psi\rightarrow -\mu^2\tau^2\psi$           \\ \hline
\end{tabular}
\captionsetup{justification=raggedright}
\caption{Lattice symmetry representations on the Dirac fermions.}
\label{table:symmetries}
\end{table}

We note that the UV time reversal $\mathcal{T}$ does not act on the $SU(3)$ orbital index.
Knowing the UV symmetry action on the partons, we now determine how $G_{UV}$ is embedded into $G_{IR}$. 
Note that, as is true in general for the projective symmetry group of spin liquids, the $PSU(3)$ symmetry of the UV lattice ansatz will be embedded as a subgroup of the $SU(6)_f$ flavor symmetry of QED$_3$.
Projected onto the IR symmetries, the lattice symmetry elements (as in Table~\ref{table:symmetries}) will also clearly include elements of $PSU(6)_f$, although now possibly supplemented by discrete QED$_3$ symmetry factors.
Fortunately, the simplicity of our ansatz allows to decompose 
\begin{equation}
\label{eq:su6branching}
    SU(6)_f\rightarrow SU(3)_s\times SU(2)_v
\end{equation} 
as all of the UV spin and lattice symmetries decouple in this manner within $SU(6)_f$.
Then, within $({(SU(6) \times U(1)_{top})})/{\mathbb{Z}_6}$, the UV symmetry operators that act on the monopoles can be thought of as elements of  
\begin{align}
  PSU(3)_s\times PSU(2)_v\times U(1)_{top}\subset \frac{(SU(6) \times U(1)_{top})}{\mathbb{Z}_6}.
\end{align} 
The $PSU(3)$ symmetry embeds into the IR symmetry $(SU(6) \times U(1)_{top})/\mathbb{Z}_6$ exactly as induced by the branching in Eq.~\eqref{eq:su6branching}, which factors the $SU(6)$ flavor symmetry into a part coming from the UV $SU(3)$ symmetry and an emergent $SU(2)$ valley symmetry.
From the action of the $G_{UV}$ symmetries (Table~\ref{table:symmetries}) on $\psi$, we can deduce how the lattice symmetries are embedded in 
\begin{equation}
  PSU(3)_s\times PSU(2)_v\times SO(2,1)_L.
\end{equation}
For example, one can see that $T_{1}$ will be embedded as the element $\mathbb{I}_{3\times 3}\otimes e^{i\vec{K}\cdot\vec{R}_1\mu^3}\otimes \mathbb{I}_{2\times 2}$, while an $SU(3)$ symmetry $g$ will be embedded as $g\otimes \mathbb{I}_{2\times 2}\otimes \mathbb{I}_{2\times 2}$.
For calculational ease, it is useful to separate the fundamental monopoles into a basis that respects the branching in Eq.~\eqref{eq:su6branching}, and this procedure is outlined in Appendix~\ref{app:monopole}.

However, there are two complications we must address.
Firstly, the embedding of the lattice symmetries $\varphi(G_{UV})$, into entire IR symmetry group
seems unclear, as the action of $G_{UV}$ on $\psi$ does not completely specify how to incorporate the IR discrete Lorentz symmetries, which from Eq.~\eqref{eq:bareaction_monopole} all act the monopole operators as some combination of Hermitian conjugation and multiplication by the invariant tensor $E_{AB}$.
In previous works that considered $N_f\equiv 0\pmod{4}$ \cite{Song_2020,calvera2021theory}, $E_{AB}$ and the other actions of the IR discrete symmetries could be included within the $SU(N)_s\times SU(M)_v$ action.
However, for our case where $N_f=6$, this is no longer true, and the IR symmetries cannot always simply be absorbed into $SU(3)_s\times SU(2)_v$ as transformations as $E_{AB}$ does not respect this branching.
Instead, the embedding of the lattice symmetries should arise from comparing Eqs.~\eqref{eq:bareaction_dirac} to \eqref{eq:bareaction_dirac_end} to the PSG action, as done in \cite{Song_2019}. For example, on the monopole operators, lattice time reversal acts as the IR time reversal on the $\psi$'s composed with a valley rotation $\mu^2$.

The second complication, as noted earlier, is the presence of a phase ambiguity associated with the $U(1)_{top}$ factor.
From $G_{UV}$, there is no information in the projective symmetry group regarding $U(1)_{top}$.
Furthermore, the $U(1)_{top}$ factor involves inherently UV physics, as it can be thought of as an effective Berry phase accumulated when the monopoles moves in a lattice charge background.
In Appendix~\ref{app:berry}, we use analytic arguments and techniques from band theory as described in \cite{Song_2019,Song_2020} to identify to derive the correct $U(1)_{top}$ Berry phase.  

Using those results, we can explicitly write down the embedding $\varphi$ that maps the UV symmetry elements into 
\begin{equation}
    \frac{(SU(6) \times U(1)_{top})}{\mathbb{Z}_6}
\end{equation}
in addition to Lorentz group factors and discrete symmetries.
The mapping is as follows:
\begin{align}
\varphi(g\in SU(3))&=(g\otimes\mathbb{I}^v_{2\times 2}\otimes 1^{top})\otimes \mathbb{I}^L_{2\times 2},
\nonumber
\\
\varphi(T_{1,2})&=\left(\mathbb{I}^s_{3\times 3}\otimes e^{i\vec{K}\cdot\vec{R}_{1,2}\mu^3}\otimes 1^{top}\right)\otimes \mathbb{I}^L_{2\times 2},
\nonumber
\\
\varphi(C_6)&=\left(\mathbb{I}^s_{3\times 3}\otimes i\mu^1\otimes 1^{top}\right)\otimes e^{-\frac{i\pi}{6}\tau^3},
\nonumber
\\
\varphi(\mathcal{R}_y)&=\left(\mathbb{I}^s_{3\times 3}\otimes i\mu^3\otimes 1^{top}\right)\otimes \mathbb{I}_{2\times 2}^L\circ \mathcal{R}_{IR},
\nonumber
\\
\varphi(\mathcal{T})&=\left(\mathbb{I}^s_{3\times 3}\otimes i\mu^2\otimes i\right)\otimes i\tau^2\circ \mathcal{T}_{IR}.
\end{align}
In the above, we have denoted the matrices $\otimes\mathbb{M}^{s,v,top,L}$ to act on the $SU(3)_s$, $SU(2)_v$, $U(1)_{top}$, and Lorentz group factors, respectively.

From our analysis in Appendices~\ref{app:monopole} and \ref{app:berry}, we find a single trivial monopole
\begin{equation}
    \operatorname{Re}(\Phi_1^{\dagger}+\Phi_2^{\dagger})+\operatorname{Im}(\Phi_1^{\dagger}+\Phi_2^{\dagger}).
\end{equation}
As mentioned earlier, we see that the $N_f=6$ charge one monopole is a relevant perturbation, leading to the Dirac spin liquid not being a stable gapless phase as the monopole proliferates to confine gauge fluctuations and a fermion mass develops to gap the matter fields.
Most likely, there will be an instability towards some proximate symmetry breaking orders.
In our case, there are also many symmetry allowed mass-monopole and monopole-monopole billinear couplings, though these are irrelevant due to their large scaling dimension.

Therefore, our analysis supports that the DSL is a quantum critical point, tuned by the proliferation of a single symmetry allowed monopole.
We will assume that the UV symmetry
allowed operators in the higher representations of
$SU(6)$ are irrelevant.
These assumptions, discussed in in Appendix~\ref{app:pairs}, are similar to
the ones needed for a stable Dirac spin liquid phase for $N_f=4$ on the kagome and triangular lattices, and a DSL critical point on the square lattice \cite{he_scipost_2022,zhang2024diracspinliquidunnecessary}.
In Appendix~\ref{app:chiral}, we also comment on the possibility that a chiral mass $\overline{\psi}\psi$ is condensed, which would suppress monopole proliferation. However, this requires spontaneous time reversal symmetry breaking, which we will not consider further.
\section{'t Hooft anomalies of $N_f=6$ QED$_3$}
\label{sec:anomalies}
In this section, we will focus on the continuum theory and perform a  systematic analysis of the quantum anomalies associated with $N_f=6$ QED$_3$, Eq.~\eqref{eq:qed_action}, with and without monopoles.
The anomalies strongly constrain the IR behavior of the (perturbed) DSL.
This has been done in other cases in \cite{Song_2020,calvera2021theory,hsin_jhep_2024,dumitrescu_arxiv_2024}, but our calculation considers the anomaly in the presence of monopole perturbations. 
Furthermore, the Lieb-Schultz-Mattis-Oshikawa-Hastings (LSMOH) theorems \cite{lieb1961,Oshikawa2000,Hastings2004} forbid some systems from having a trivially gapped state and can be used in conjunction with the 't Hooft anomalies
in order to restrict the possible values of monopole phases.
However, the LSMOH anomalies associated with translation vanish when there are $N$ fundamental $SU(N)_s$ spins per unit cell, which is true for our kagome lattice ansatz.
However, there is an anomaly between the $SU(3)_s$ symmetry and rotation (or reflection), which we will derive from the IR anomaly.

The 't Hooft anomaly of the symmetry $G$ can be probed by attempting to gauge $G$.
The action that arises can be interpreted as living on the boundary of a $(3+1)d$ symmetry protected topological phase on a manifold $M_4$, and the anomaly can be detected by dependence of the partition function on the choice of $M_4$ or the extension of gauge fields to $M_4$.
We will consider two relevant choices of $G$:
\begin{itemize}
    \item $(SU(6)\times U(1)_{top})/\mathbb{Z}_6$,
    the internal symmetry of the QED$_3$ fixed point. 
    \item $(SU(3)\times SU(3))/\mathbb{Z}_3$, corresponding to the kagome DSL case, in which we have perturbed the CFT with  the symmetry-allowed monopole.
    The quotient $\mathbb{Z}_3$ is the diagonal center of each $SU(3)$ factor.
\end{itemize}
(We will also briefly consider a ``mid-IR" theory of Dirac fermions coupled to a $U(1)$ gauge field with $PSU(3)_s\times PSU(2)_v\times U(1)_{top}$, relevant to Dirac spin liquids with $SU(3)$ spin and $SU(2)$ valley symmetries.)
In all cases, we find a nontrivial 't Hooft anomaly, precluding the existence a trivial, symmetric, gapped ground state.
Furthermore, we argue that the form of the anomaly with the monopoles included prohibits a symmetric topologically ordered phase, which is an example of symmetry-enforced gaplessness \cite{wang_prb_2014,wang2016composite,sodemann2017composite,wang2017deconfined,cordova2019anomaly,cordova2020anomaly}.
Therefore, the only allowed IR theory in the presence of the monopole must be one that is symmetry breaking or, if symmetry preserving,  must be gapless.
\subsection{$(SU(6)\times U(1)_{top})/\mathbb{Z}_6$}
In this section, we will explore the anomalies of QED$_3$ without monopoles, but with the full IR $(SU(6)\times U(1)_{top})/\mathbb{Z}_6$ symmetry.
We will also introduce many tools and methods that will be useful in analyzing the later cases with different symmetry.

It will be convenient to place the continuum theory on an arbitrary closed oriented 3-manifold $M_3$ with a metric $g$ and to introduce background gauge fields for the $(SU(6) \times U(1)_{top})/\mathbb{Z}_6$ global internal symmetry. 
In the presence of these background gauge fields and metric, the action $S$ will not be well-defined as a $(2+1)d$ theory; rather it requires an extension of $g$ and the background gauge fields (though not the dynamical fields) to a bulk $(3+1)d$ theory. 
We can view the action as arising from a $(3+1)d$ SPT bulk on a manifold $M_4$ action whose restriction to the boundary $M_3 = \partial M_4$ yields the exact gauged action for our original $N_f=6$ QED$_3$ theory.
To characterize the anomaly, it is sufficient to consider the bulk theory on a closed manifold $M_4$:  the difference in the partition function phase between two choices of an open $M_4$ could be detected through $S_{bulk}$
evaluated on the closed manifold $\tilde{M}_4$, where $\tilde{M}_4$ is obtained by gluing the two choices of $M_4$ along their boundaries.

We thus turn on background gauge fields  $\mathcal{A}^6$  for $SU(6)$ and $\mathcal{A}^{top}$  for $U(1)_{top}$. Due to the $\mathbb{Z}_6$ quotient, all local operators in the theory transform faithfully under $PSU(6)$, and we will therefore regard $\mathcal{A}^6$ as a $PSU(6)$ gauge bundle. Not every such bundle can be lifted to an $SU(6)$ bundle. The obstruction to doing so is captured by a characteristic class \cite{duncan2013componentsgaugegrouppurbundles,Kapustin2014coupling,Genolini2021}, known as the Brauer or Stiefel-Whitney class, $w_2 \in H^2(M_4, \mathbb{Z}_6)$. 

To ensure that both $SU(6)$ and $U(1)_{top}$ have a common quotient by $\mathbb{Z}_6$, we also have the cocycle condition
\begin{equation}
\label{eq:daTOP_cocycle}
    \oint \frac{1}{2}w_2+\frac{d\mathcal{A}^{top}}{2\pi}\in\mathbb{Z}.
\end{equation}
To see how this arises, note that for each $2\pi$ strength monopole, there are six zero modes, of which three of them will attach to and dress the bare monopole.
Rotating by an element $e^{\frac{2\pi i m}{6}}\mathbb{I}$ in the $\mathbb{Z}_6$ center of the flavor $SU(6)$ contributes to a phase of $e^{\frac{2\pi i m}{2}}$, which can be compensated by a $U(1)_{top}$ rotation.
Therefore, only $m\pmod{2}$ is important and can lead to a fractional flux of $A_{top}$.

In what follows we will construct the QED$_3$ theory by first considering a `parent' theory of $N_f = 6$ massless Dirac fermions in $(2+1)d$  and then coupling in the dynamical $U(1)$ gauge field. The parent Dirac fermion theory has a global $U(6) = ({U(1)_a \times SU(6)})/{\mathbb{Z}_6}$ symmetry and we include a coupling to a background $U(6)$ gauge field. We have denoted the $U(1)$ factor in $U(6)$ as $U(1)_a$ and the corresponding gauge field will be denoted $a$. For now $a$ is a background gauge field but we will soon promote it to a dynamical gauge field to define the QED$_3$ theory. 

In the parent theory, the only fields with unit charge under $U(1)_a$ transform in the fundamental of $SU(6)$ and are fermions (spinors of the tangent bundle of $M_4$). 
Hence we obtain the twisted flux condition
\begin{equation}
\label{eq:spinc_cocycle}
    \oint \frac{1}{6}w_2+\frac{1}{2}w_2^{TM}+\frac{da}{2\pi}\in\mathbb{Z},
\end{equation} 
which ensures that the $U(6)$ gauge field couples properly to fermions.
In the above, we have used $w_2^{TM}\in H^2(M_4,\mathbb{Z}_2)$ to be the Stiefel–Whitney class of the tangent bundle of $M_4$.
Therefore, in the presence of $SU(6)$, the $U(1)$ gauge field $a$ becomes a generalized form of a spin$_c$ connection.
Intuitively, this relation captures the twisting of the transition functions that are allowed because in order for the fermions to be well-defined, a particular combination of the gauge bundle transition functions need to satisfy the usual cocyle compatibility condition. More physically, a defect in the cocycle condition of one principal gauge bundle can be compensated for by some nontrivial flux in another bundle.

Now to explicitly derive the action, we will Pauli-Villars regularize the Dirac fermions, where we define the partition function of a Dirac fermion coupled to a general gauge field $A$ and gravitational metric $g$ to be
\begin{equation}
    Z[A,g]_{PV}=|Z[A,g]|\exp\left(-\frac{i\pi}{2}\eta[A,g]\right).
\end{equation}
Here $\eta[A,g]$ is the $\eta$ invariant of the Dirac operator \cite{witten2016fermion,seiberg2016duality,senthil2019duality}, which is classically equal to the half-integer level Chern-Simons (CS) term arising from an additional gapped Dirac fermion. 
To define the Dirac theory, we must extend $g$, $\mathcal{A}_{top}$, and the $U(6)$ gauge field $\hat{\mathcal{A}}(a,\mathcal{A}^6)$ into the bulk manifold $M_4$. 
We now write the free $N_f = 6$ Dirac fermion theory, together with its background gauge fields as
\begin{equation}
    Z_{Dirac}=\int \left[\mathcal{D}\psi\right]e^{-S_{boundary}[\psi,a,\mathcal{A}^6,\mathcal{A}^{top},g]-S_{bulk}[a,\mathcal{A}^6,\mathcal{A}^{top},g]}.
\end{equation} 
From the $\eta$ invariant, one can find
\begin{widetext}  
\begin{equation}
\label{eq:gauged_psu6}
    S=S_{boundary}+S_{bulk}=\int \left[\left(\overline{\psi}\slashed{D}_{[a,\mathcal{A}^6,\mathcal{A}^{top},g]}\psi\right)_{PV}+i\frac{3}{2}CS[a]+i\frac{1}{2}CS[\mathcal{A}^6]+6iCS_{grav}[g]-i\frac{1}{2\pi}a\wedge d\mathcal{A}^{top}\right],
\end{equation}
\end{widetext}
where the additional topological terms are introduced to properly preserve time reversal $\mathcal{T}$ in the presence of the the $\eta$ invariant.
The CS terms are defined by 
$dCS[A]=\frac{1}{4\pi}\tr_{fund}[F\wedge F]$ and $dCS[g]=\frac{1}{196\pi}\tr_{vec}[R\wedge R]$,
where $F=dA+A\wedge A$ is the curvature of $A$ and $R$ is the Riemann curvature tensor of $M_4$.

As discussed earlier, half-integer CS levels are not gauge invariant, so it is important to view them as arising from a $(3+1)d$  bulk action on a manifold $M_4$ action whose restriction to the boundary $\partial M_4$ yields exact gauged action $S$ in Eq.~\eqref{eq:gauged_psu6}.
Using Stokes' theorem, we can write this bulk action as
\begin{equation}
\label{eq:psu6_anomaly_first}
    \frac{S_{bulk}}{2\pi i}=\frac{3}{2}p_1[a]-\frac{3}{8}\sigma+\frac{1}{2}p_1[\mathcal{A}^6]-\int_{M_4}\frac{da}{2\pi}\wedge\frac{d\mathcal{A}^{top}}{2\pi},
\end{equation}
where $p_1[A]$ is the instanton/Pontryagin number of the gauge field $A$, defined by 
\begin{equation}
p_1[A]=\frac{1}{8\pi^2}\int_{M_4}\tr_{Fund}[F_A\wedge F_A].
\end{equation}
In the case of $U(1)$, we define $p_1[a]=\frac{1}{4\pi^2}\int_{M_4}da\wedge da$.
Extending the gravitational CS term into the bulk yields the signature term in the bulk action
\begin{equation}
    \sigma=-\frac{1}{24\pi^2}\int_{M_4}\tr_{vec}[R\wedge R].
\end{equation}

With this definition of the parent free Dirac theory coupled to background gauge fields, we can promote $a$ to be dynamical. It is important however that once thus promoted the bulk action does not depend on $a$ but only on the remaining background gauge fields.
Thus we will first eliminate the dependence of the bulk action above on $a$, and express it entirely in terms of the remaining true background gauge fields.  

Let us focus on the first two anomaly terms in Eq.~\eqref{eq:psu6_anomaly_first}.
Note that if $a$ was a genuine spin$_c$ connection, then $p_1[a]/2-\sigma/8\in\mathbb{Z}$, so that there will be no contribution from that part of the anomaly to the bulk action as 
\begin{equation}
Z_{Dirac}\sim \exp\left[2\pi i\left(\frac{3p_1[a]}{2}-\frac{3\sigma}{8}\right)\right]=1,
\end{equation}
independent of the choice of extension into $M_4$.
However, $a$ is twisted by the cocycle condition, Eq.~\eqref{eq:spinc_cocycle}.
Consequently, unless the Brauer class of $\mathcal{A}^6$ vanishes,
\begin{equation}
    w_2\in H^2(M_4,\mathbb{Z}_6)=0,
\end{equation}
$p_1[a]/2-\sigma/8$ will not be in $\mathbb{Z}$.
Therefore, as expected the anomaly does not depend on $a$ and we can hence express the bulk action involving $a, \mathcal{A}^6, g, A_{top}$ entirely in terms of $(\mathcal{A}^6,g, A_{top})$. 
To that end, let $C$ be a genuine spin$_c$ connection \footnote{Note that every oriented 4-manifold is also a spin$_c$ manifold. This guarantees that there always exists such a spin$_c$ connection $C$.}  satisfying
\begin{equation}
\label{eq:dC_cocycle}
    \oint\dfrac{dC}{2\pi}=\oint \frac{w_2^{TM}}{2}\pmod{1}.
\end{equation}
Then we have $B=a-C$ a representative $U(1)$ gauge field with shifted periods by $w_2/6$.
We can subsequently simplify
\begin{equation}
    \begin{aligned}
    p_1[a]= p_1[B+C]=p_1[B]+p_1[C]+2\int
    \dfrac{dC}{2\pi}\wedge \dfrac{dB}{2\pi}.
\end{aligned}
\end{equation}
Using that 
\begin{equation}
    3\left(\dfrac{p_1[C]}{2\pi}-\dfrac{\sigma}{8}\right)\in\mathbb{Z},
\end{equation}
we can now simplify the first two terms in Eq.~\eqref{eq:psu6_anomaly_first} to
\begin{equation}
    \frac{S_{bulk}}{2\pi i}=\dfrac{3}{2}p_1[B]+3\int
    \dfrac{dC}{2\pi}\wedge \dfrac{dB}{2\pi}+\cdots
\end{equation}
Now we simplify the last term in Eq.~\eqref{eq:psu6_anomaly_first} by writing
\begin{align}
    \int \frac{da}{2\pi}\wedge\frac{d\mathcal{A}^{top}}{2\pi}&=\int \frac{dB+dC}{2\pi}\wedge\frac{d\mathcal{A}^{top}}{2\pi}.
\end{align}
As $C$ is a spin$_c$ connection, we can rewrite
\begin{widetext}
\begin{align}
\int \frac{dC}{2\pi}\wedge\frac{d\mathcal{A}^{top}}{2\pi}&=3\int \frac{dC}{2\pi}\wedge\frac{dB}{2\pi}+\int \frac{dC}{2\pi}\wedge\frac{d(\mathcal{A}^{top}-3B)}{2\pi}\\
&=3\int \frac{dC}{2\pi}\wedge\frac{dB}{2\pi}+\frac{1}{2}\left(p_1[\mathcal{A}^{top}-3B+C]-p_1[C]\right)-\int\frac{d(\mathcal{A}^{top}-3B)}{2\pi}\wedge\frac{1}{2}\frac{d(\mathcal{A}^{top}-3B)}{2\pi}\\
&=3\int \frac{dC}{2\pi}\wedge\frac{dB}{2\pi}-\frac{1}{2}\int\frac{d(\mathcal{A}^{top}-3B)}{2\pi}\wedge\frac{d(\mathcal{A}^{top}-3B)}{2\pi}
\end{align}
Combining the above and substituting into Eq.~\eqref{eq:psu6_anomaly_first}, we obtain 

    \begin{equation}
    \label{eq:sbulk_prefinal}
    \frac{S_{bulk}}{2\pi i}=\frac{1}{2}p_1[\mathcal{A}^6]+6p_1[B]+\int\dfrac{d\mathcal{A}^{top}}{2\pi}\wedge\left(\frac{1}{2}\dfrac{d\mathcal{A}^{top}}{2\pi}+4\cdot\dfrac{dB}{2\pi}\right).
\end{equation}
We now observe that
\begin{align}
    2\dfrac{d\mathcal{A}^{top}}{2\pi}\wedge\dfrac{dB}{2\pi}&=\dfrac{d\mathcal{A}^{top}}{2\pi}\cup\frac{w_2}{3}\pmod{1},\\
    6p_1[B]&=\frac{1}{6}\mathcal{P}(w_2)\pmod{1},
\end{align}
where we have used the Pontryagin square operation \cite{whitehead1949simply,kapustin2013topologicalfieldtheorylattice}, $\mathcal{P}\ :\ H^2(M_4,\mathbb{Z}_{2n})\rightarrow H^4(M_4,\mathbb{Z}_{4n})$.
Therefore, we can obtain the final form of the anomaly
\begin{equation}
\label{eq:final_anomaly_psu6}
    \frac{S_{bulk}}{2\pi i}=\frac{1}{2}p_1[\mathcal{A}^6]+\frac{1}{6}\int\mathcal{P}(w_2)+\int \frac{d\mathcal{A}^{top}}{2\pi}\cup\left(\frac{1}{2}\frac{d\mathcal{A}^{top}}{2\pi}+\frac{2w_2}{3}\right)\pmod{1}.
\end{equation}
\end{widetext}
Note that all dependence of the bulk topological action on $a$ has been eliminated, and thus we can safely promote $a$ to a dynamical gauge field in the boundary $(2+1)d$ theory. Eqn. \ref{eq:final_anomaly_psu6} and the subsequent re-expression below (Eqn. \ref{eq:final_anomaly_psu6_p1}) are thus the bulk anomaly theory capturing the anomaly of the IR global symmetry of the $N_f = 6$ QED$_3$ theory. 

The first term Eqn. \ref{eq:final_anomaly_psu6} is half of the Pontryagin number of $\mathcal{A}^6$, which is a $\theta=\pi$ term for the $SU(6)$ bundle described by $\mathcal{A}^6$. 
Because $\mathcal{A}^6$ is twisted by the center $\mathbb{Z}_6$ symmetry (or equivalently, it has a nonzero Brauer class $w_2$), $p_1[\mathcal{A}^6]$ has a fractional part given by $\frac{5}{12}\int\mathcal{P}(w_2)$ \cite{woodward_lsm_1982,duncan2013componentsgaugegrouppurbundles}. Writing $p_1[\mathcal{A}^6]$ as \begin{equation}
    p_1[\mathcal{A}^6]=\mathcal{I}[\mathcal{A}^6]+\frac{5}{12}\int\mathcal{P}(w_2),\quad \mathcal{I}[\mathcal{A}^6]\in\mathbb{Z},
\end{equation}
we do not have a well defined expression for $\frac{1}{2}p_1[\mathcal{A}^6]$ purely in terms of the Brauer class $w_2$, as $\frac{1}{2}p_1[\mathcal{A}^6]$ depends on not only the fractional part of the instanton number, but also the integer part modulo $2$. Note that while the relation between instanton number and $w_2$ holds generally, it can be seen
explicitly in the special case where there exists a lift to a $U(6)$ bundle.
Then, we can write
\begin{equation}
    \mathcal{A}^6=\mathcal{A}_{U(6)}-B\mathbb{I}_{6\times 6},\quad \oint\dfrac{dB}{2\pi}=\oint \frac{w_2}{6}\pmod{1}
\end{equation}
in terms of a $U(6)$ gauge field $\mathcal{A}_{U(6)}$ under the restriction $\tr[\mathcal{A}_{U(6)}]=6B$. One can then calculate that 
\begin{equation}
    p_1[\mathcal{A}^6]=15\int \dfrac{dB}{2\pi}\wedge\dfrac{dB}{2\pi}=\dfrac{5}{12}\int\mathcal{P}(w_2)\pmod{1}
\end{equation}

Going back to Eq.~\eqref{eq:final_anomaly_psu6}, we see there is a flavor symmetry anomaly arising from $\frac{1}{6}\int \mathcal{P}(w_2)$.
This contribution is nothing more than the fractional part of a $\theta=-4\pi$ term for the $SU(6)$ gauge field $\mathcal{A}^6$. Equivalently, one can see
$-2p_1[\mathcal{A}^6]=\frac{1}{6}\int \mathcal{P}(w_2)\pmod{1}$, so we can actually rewrite the final form of the anomaly, Eq.~\eqref{eq:final_anomaly_psu6}, as
\begin{widetext}
\begin{align}
    \label{eq:final_anomaly_psu6_p1}
        \frac{S_{bulk}}{2\pi i}=-\frac{3}{2}p_1[\mathcal{A}^6]+\int \frac{d\mathcal{A}^{top}}{2\pi}\cup\left(\frac{1}{2}\frac{d\mathcal{A}^{top}}{2\pi}+\frac{2w_2}{3}\right)\pmod{1},
\end{align}
\end{widetext}
so that the flavor symmetry is completely captured by a $\theta=-3\pi$ term for the $SU(6)$ gauge field. We note here the presence of the quotiented $\mathbb{Z}_6$ center symmetry makes $\theta$ periodic under $\theta\sim \theta+24\pi$.

The last term in Eq.~\eqref{eq:final_anomaly_psu6_p1} has a simple interpretation: namely, magnetic particles of $U(1)_{top}$ transform in a
representation of $SU(6)$ with $N$-ality $2$ and  are fermions.
Furthermore, the $\theta$ term for $U(1)_{top}$ ensures that magnetic particles of $U(1)_{top}$ carry a single electric charge of $U(1)_{top}$.
Fusing these magnetic particles with the original local monopole, one obtains a particle with zero $U(1)_{top}$ electric charge, a unit $U(1)_{top}$ magnetic charge, and $1$-ality under $SU(6)$; this particle can be interpreted to be the remnant of the original Dirac fermion.
From another perspective, the bare monopole under $\mathcal{A}^{top}$ carries charge under the dynamical $U(1)$ field $a$, it
can be made local by dressing it with $\psi$, imparting quantum numbers of $\psi$ to the ${\mathcal{A}}^{top}$
monopole.
We also remark that $\frac{1}{2}\frac{d{\mathcal{A}}^{top}}{2\pi}\cup\frac{d{\mathcal{A}}^{top}}{2\pi}$ is the bulk theory for the nontrivial bosonic topological insulator \cite{vishwanath_prx_2013}, in which the monopoles are electrically neutral under $d{\mathcal{A}}^{top}$ and fermionic, as is reproduced here.

Note that Eq.~\eqref{eq:final_anomaly_psu6_p1} also describes a time reversal invariant theory.
This can be derived from using that $\mathcal{A}^{top}$ and the original $U(1)$ gauge field $a$ (and equivalently, $B$) have opposite transformations under time reversal.
Then, under time reversal, the first two terms in Eq.~\eqref{eq:final_anomaly_psu6_p1} change sign,
from which we can confirm
\begin{widetext}
\begin{align}
    \frac{\mathcal{T}(S_{bulk})-S_{bulk}}{2\pi i}
    &=3p_1[\mathcal{A}^6]-\int \frac{d\mathcal{A}^{top}}{2\pi}\cup\frac{d\mathcal{A}^{top}}{2\pi}=0\pmod{1}.
\end{align}
\end{widetext}
As desired, the final bulk theory is time reversal invariant on a closed manifold.
However, in the presence of a boundary, there are mixed anomalies between time reversal and the flavor symmetry and between time reversal and $U(1)_{top}$.
This can be seen from analyzing the first two terms in Eq.~\eqref{eq:final_anomaly_psu6_p1}, in which restricting the action on the boundary of an open manifold leads to half quantized Chern Simons terms for both the flavor and $U(1)_{top}$ gauge fields.
On the boundary, time reversal shifts the action by an odd integer Chern Simons term for both the $SU(6)$ and $U(1)_{top}$ gauge fields. The anomaly is nontrivial as there is no local counterterm one can add to the original theory to eliminate this shift.

From the nontrivial anomalies, we see that the ultimate fate of $N_f=6$ QED$_3$ must be a gapless phase (such as a CFT), a symmetry broken phase, or a symmetry-preserving gapped phase with topological order.
For $N_f=2$, symmetry-enforced gaplessness is also present, meaning no gapped symmetric topological order can saturate the anomaly.
Instead, symmetry breaking is the most likely fate \cite{dumitrescu_arxiv_2024} for the IR theory as bootstrap \cite{li2022plb} does not support a stable CFT.

For the case of $N_f=6$ and the general case of pure QED$_3$ with $N_f\ge 4$, we could not find an analytic argument precluding a gapped, symmetric topologically ordered state in the IR. 
However, bootstrap studies \cite{chesterjhep2016,li_2022_jhep,poland_prd_2022,he_scipost_2022} are consistent with the assumption QED$_3$ with $N_f\ge 4$ realizes a CFT in the IR.
Lastly, we remark that there is always an allowed chiral topological order in the IR. As was argued in the previous section, such a state can be realized by the condensation of a chiral mass.

In Appendix~\ref{app:anom_midIR}, we describe the anomaly in the case we take the global symmetry to be $PSU(3)_s\times PSU(2)_v\times U(1)_{top}$, which is relevant as a mid-IR theory of lattice Dirac spin liquids.

\subsection{$(SU(3)\times SU(3))/\mathbb{Z}_3$}
Lastly, we consider including  the single symmetry-allowed relevant monopole.
Recall the monopole is of the form
$(\Phi_1^{\dagger}+\Phi_2^{\dagger})\sim (f_{[1}^\dagger f_{2}^{\dagger} f_{3]}^{\dagger}\mathcal{M}^{\dagger}+f_{[4}^\dagger f_{5}^{\dagger} f_{6]}^{\dagger}\mathcal{M}^{\dagger})$.
In this case, we have broken the $U(1)_{top}$ symmetry.
However, we have still preserved a $(SU(3)\times SU(3))/\mathbb{Z}_3$ symmetry, where each factor acts as the fundamental on $(f_1,f_2,f_3)$ and 
$(f_4,f_5,f_6)$, respectively, and the common center is quotiented out.
There are also discrete residual symmetries, including a discrete $\mathbb{Z}_2$ that acts as swapping $f_{1,2,3,4,5,6}\rightarrow -if_{4,5,6,1,2,3}$ on the fermions combined with a $U(1)_{top}$ $\mathcal{M}^{\dagger}\rightarrow i\mathcal{M}^{\dagger}$, which acts as $\Phi_1\rightarrow \Phi_2$ on the monopole. In the UV, this $\mathbb{Z}_2$ representings the $C_6$ symmetry swapping the two Dirac valleys.

We will probe the $(SU(3)\times SU(3))/\mathbb{Z}_3$ response by coupling in  $SU(3)$ gauge fields $\mathcal{A}$ and $\mathcal{B}$.
Note the discrete $\mathbb{Z}_2$ exchanges $\mathcal{A}$ and $\mathcal{B}$.
In the UV, $\mathcal{A}$ and $\mathcal{B}$ represent gauge fields coupling to the two Dirac valleys.

We observe that the separation of the original flavor symmetry into a block diagonal $(SU(3)\times SU(3))/\mathbb{Z}_3$ form constrains the $\mathbb{Z}_3$ Brauer class of each block to be equal. 
In other words,
defining the classes $w_2^{\mathcal{A}},w_2^{\mathcal{B}}\in H^2(M_4,\mathbb{Z}_3)$, we have that $w_2^{\mathcal{A}}=w_2^{\mathcal{B}}\equiv u_2$.
Therefore, one can equivalently think of $(\mathcal{A},\mathcal{B})$ as two $PSU(3)$ gauge fields that are constrained to have the same obstruction class. We see that upon breaking the $PSU(6)\rightarrow (SU(3)\times SU(3))/\mathbb{Z}_3$, the Brauer class $w_2\in H^2(M,\mathbb{Z}_6)$ of the parent QED$_3$ is reduced to the diagonal obstruction class $u_2\in H^2(M,\mathbb{Z}_3)$ by the relationship $w_2=2 u_2\pmod{6}$.

The anomaly for the theory with the monopole included can be calculated with ease using the result in Eq.~\eqref{eq:final_anomaly_psu6_p1} found for pure QED$_3$ and breaking the global symmetry accordingly.
As the monopole has an expectation value, the $U(1)_{top}$ symmetry is explicitly broken, and hence it is not meaningful to turn on a background gauge field  $\mathcal{A}^{top}$.
Furthermore, the $PSU(6)$ Pontryagin number is changed from $p_1[\mathcal{A}^6]\rightarrow p_1[\mathcal{A}]+p_1[\mathcal{B}]$.
We then obtain 
    \begin{equation}
    \label{eq:anom_mono_cond}
    \frac{S_{bulk}}{2\pi i}=-\frac{3}{2}\left(p_1[\mathcal{A}]+p_1[\mathcal{B}]\right).
\end{equation}
The above represents a $\theta_{\mathcal{A},\mathcal{B}}=-3\pi$ term for both $SU(3)$ gauge fields $\mathcal{A}$ and $\mathcal{B}$.
Note that under ${\cal T}_{IR}$, both $\mathcal{A}$ and $\mathcal{B}$ change sign, and thus $\theta_{\mathcal{A},\mathcal{B}} \rightarrow - \theta_{\mathcal{A},\mathcal{B}}$. Because the fractional part of $p_1[\mathcal{A}]$ and $p_1[\mathcal{B}]$ are both the same, being $\frac{1}{3}\int\mathcal{P}(w_2)$, we see the above is $\mathcal{T}_{IR}$ invariant at any $\theta$ a multiple of $3\pi$. 
Similarly since the unitary $\mathbb{Z}_2$ symmetry exchanges $\mathcal{A}$ and $\mathcal{B}$, it follows that the anomaly theory is invariant under this symmetry so long as $\theta_{\mathcal{A}} - \theta_{\mathcal{B}} = 0$ modulo $6\pi$. As each of $\theta_{\mathcal{A}, \mathcal{B}}$ is separately $6\pi$ periodic, it follows that the requirement of $\mathbb{Z}_2$ and $\mathcal{T}_{IR}$ invariance constrains $\theta_{\mathcal{A}} = \theta_{\mathcal{B}}$ to both either be $0$ or to both  be $3\pi$ (modulo $6\pi$). The specific theory we have realizes the second possibility where the corresponding boundary theory then has a mixed anomaly between $(SU(3)\times SU(3))/\mathbb{Z}_3$ and $\mathcal{T}_{IR}$. 
This implies that the field theory cannot realize a gapped state symmetric under the $(SU(3)\times SU(3))/\mathbb{Z}_3$,  ${\cal T}_{IR}$, and $\mathbb{Z}_2$ symmetries. 
However, we recall that the UV time reversal $\mathcal{T}_{UV}$ actually exchanges the two Dirac valleys.
Consequently, we observe there is no mixed anomaly between the IR symmetry and $\mathcal{T}_{UV}$, which is a combination of the discrete $\mathbb{Z}_2$ exchanging the valleys and the IR time reversal.
Instead, $\mathcal{T}_{UV}$ constrains $\theta_{\mathcal{A},\mathcal{B}}=-\theta_{\mathcal{B},\mathcal{A}}$, under which the $\theta$ angles can flow to be trivial.

Though a trivial symmetry-preserving gapped state is not possible for the boundary theory, we may consider if the IR theory allows a symmetric gapped topological order. We will argue that such a state is not allowed. To see this, can focus on the first $SU(3)$ factor and break it into its maximal torus, $U(1)^2$. Threading a monopole through its first $U(1)$ subgroup (with normalization such that the $SU(3)$ fundamental carries unit electric charge), we see that the anomaly enforces that the monopole has electric charge $(1,1/2)$. Therefore, for $U(1)^2$ charge to be conserved, there must be a quasiparticle in this surface state with these properties. However, the only such particles are those that transform linearly under $SU(3)$, with integer charges under the $U(1)^2$. 
Therefore, the hypothetical gapped state cannot saturate the anomaly \footnote{If time reversal symmetry is broken, such as if we allow a chiral mass $m\overline{\psi}\psi$, there can be a surface Hall conductivity for the $SU(3)$.
This leads to the monopole threading event to nucleate a $(1,1/2)$ charge on the surface as required, and there is no obstruction to a gapped symmetry preserving state.}.

Accordingly, we see the ultimate fate of QED$_3$ perturbed by the allowed monopole must be a symmetry breaking but otherwise trivial phase (where the $(SU(3) \times SU(3))/\mathbb{Z}_3 \times \mathbb{Z}_2$ is broken), an exotic conformal fixed point, or a time reversal broken topologically ordered phase. While the latter two cannot be ruled out, we conjecture that the simplest possible symmetry option, namely, the symmetry broken phase is realized. We will simply assume that this is the case in our subsequent discussion. 
\subsection{Unnecessary Criticality}
As the anomaly of the DSL with monopole is $\mathbb{Z}_2$ classified, it follows that the anomaly is the same for either sign of $\lambda$.
Recall that given the trivial monopole $\Phi^{\dagger}_{triv}$,
the phase transition is described by
\begin{equation}
    \mathcal{L}_{DSL}+\lambda(\Phi^{\dagger}_{triv}+h.c.)+\cdots,
\end{equation}
where $\cdots$ includes (irrelevant) terms allowed by the microscopic symmetries.
We now argue that the DSL is an ``unnecessary'' quantum critical point, which lies entirely within a  single phase \cite{bi_prx_2019}.

Note that the two signs of $\lambda$ are related by an $SU(6)$ transformation that acts on the Dirac fermions by 
\begin{equation}
\label{eq: su6lambda}
    U=\begin{pmatrix}
    0&1&0&&&\\
    1&0&0&&0_{3\times 3}&\\
    0&0&1&&&\\
    &&&0&1&0\\
    &0_{3\times 3}&&1&0&0\\
    &&&0&0&1
\end{pmatrix},\quad\psi\rightarrow U\psi.
\end{equation}
Thus the ground state of the continuum field theory for $\lambda < 0$ will be related to the ground state at $\lambda > 0$ by this $SU(6)$ transformation. 

From the previous section, we accept the conjecture that the RG flow of $\lambda$ away from the $\lambda = 0$ DSL fixed point takes the system to  a symmetry broken but otherwise trivial phase. A specific possibility is that the symmetry broken state is just the trimerized VBS ground state of the nearest neighbor spin model in Fig.~\ref{fig:vbs}. Microscopically this state preserves the $PSU(3)$, translation, and $\mathcal{T}_{UV}$ symmetries but breaks $C_6$ rotation. In the IR theory, this manifests as a state that breaks the unitary $\mathbb{Z}_2$ while preserving $(SU(3) \times SU(3))/\mathbb{Z}_3$ (and $\mathcal{T}_{UV}$). The breaking of $\mathbb{Z}_2$ trivializes the anomaly so that there need not be any further topological order or gapless excitations in the resulting ground state. Let us consider the possibility that the flow of $\lambda$ lands us in this $\mathbb{Z}_2$ broken phase. 

In terms of the Dirac fermion representation of the continuum Lagrangian, the $\mathbb{Z}_2$ symmetry breaking order parameter is $\overline{\psi} \mu^3 \psi$ 
which is invariant under the $SU(6)$ rotation in Eqn. \ref{eq: su6lambda}. It follows that if the ground state for $\lambda > 0$ breaks $\mathbb{Z}_2$ but is otherwise trivial and gapped, then so will the ground state at $\lambda < 0$. 
Then the DSL appears as an unnecessary quantum critical point inside this phase. 

Since in the original microscopic system, the trivial $\mathbb{Z}_2$ broken phase is identified with the VBS phase, we conclude that by tuning one parameter entirely inside this VBS phase, we can reach  the DSL as a critical point. Like in other examples of unnecessary criticality, in parameter space, the DSL will live on a surface of codimension one that terminates \footnote{The termination will be described by a multicritical fixed point that we will not attempt to describe here.} within the VBS phase so that there is a smooth path that connects either side of this surface. 


Alternatively, if instead $\mathbb{Z}_2$ is preserved but the $(SU(3)\times SU(3))/\mathbb{Z}_3$ symmetry is broken to some subgroup  $H$ in the $\lambda \neq 0$ continuum theory, the $SU(6)$ rotation that relates $\lambda$ to $-\lambda$ will only move us within the corresponding order parameter manifold. The precise ``orientation'' of the order parameter within this manifold is anyway not fixed by the theory itself, and hence we have the same symmetry breaking pattern for either sign of $\lambda$. Thus once again in the continuum theory we have an unnecssary critical point. In the microscopic system, the corresponding state is an insulator that breaks a UV symmetry such as $PSU(3)_s$ or lattice translation symmetry. 
Therefore, the DSL can also be an unnecessary critical point within a single $SU(3)_s$ or translation symmetry breaking phase.

\section{Discussion and Outlook}
\label{sec:conclusion}
 
In this paper, we studied the possible stability of a DSL  in  a kagome lattice   $SU(3)$ magnet.
Through a careful analysis of the quantum numbers of allowed relevant perturbations at the DSL fixed point, we showed that the DSL is not a stable phase, but instead a quantum critical point, tuned by a symmetry-allowed single-strength monopole.

We performed a detailed analysis of the emergent symmetries and 't Hooft anomalies of both the critical fixed point as well as the theory obtained in the presence of the relevant perturbation.
The presence of anomalies forbid a trivial, symmetric, gapped ground state even when the DSL theory is perturbed away from its critical fixed point. The anomaly of the resulting theory is the same for either sign of the relevant perturbation. The most likely fate is that the RG flow induced by this perturbation leads to a symmetry broken ground state. A specific possibility is a trimer valence bond solid phase that preserves translation and $SU(3)$ spin symmetries but breaks lattice $C_6$ rotations. We argued that the DSL is an unnecessary quantum critical point that 
does not separate two distinct phases but instead lies within a single phase. 
Thus if the symmetry breaking state is the trimer VBS 
state, then the DSL is a critical spin liquid living entirely inside this phase and which can be accessed by tuning one parameter. 
Along with recent results showing the DSL is also an unnecessary quantum critical point on the square lattice antiferromagnet \cite{zhang2024diracspinliquidunnecessary}, our paper suggests that unnecessary quantum criticality  may not be an uncommon phenomenon 
in quantum many body systems. 

It would be insightful for both the study of quantum field theory and spin liquids in real materials to further explore the mechanisms behind quantum critical behavior in QED$_3$ theories. Our work adds to a growing body of literature on the rich behavior of QED$_3$ in lattice systems.

The precise nature of the phase driven by the relevant symmetry-allowed monopole remains an open question, though we presented arguments in favor of a symmetry broken phase.
Numerical simulations could help settle this question and to determine the precise pattern of the presumed symmetry breaking.

Lastly, we remark that 
with recent advances in ultracold atoms experiments, there is hope that the DSL and its neighboring phases could be probed in a laboratory setting. In general, the engineering of $SU(N)$-symmetric systems allows for a rich playground on which to explore these exotic quantum phenomena and offers an exciting pathway to probe QED$_3$ in the real world.

\section{Acknowledgments}
We thank Sal Pace and Po-Shen Hsin for useful discussions. YZ was supported by the National Science Foundation Graduate Research Fellowship
under Grant No. 2141064. XYS was supported by the Croucher innovation award. TS was supported by NSF grant DMR-2206305.  

\appendix
\section{Mean field state and continuum limit}\label{app:mft}
Fourier transforming to momentum space
\begin{equation}
f_{i\alpha} (\vec{k}) = \frac{1}{\sqrt{N}} \sum_{\vec{r}\in \Lambda_i} e^{-i \vec{k}\cdot\vec{r}}f_{\vec{r}\alpha},
\end{equation}
where $N$ is the number of unit cells and $i\in\{1,2,3\}$ labels the sublattice, we obtain a mean field Hamiltonian
\begin{alignat}{1}
H_{MF} &= \sum_{\vec{k},\alpha}\mathfrak{f}^\dagger_{\alpha}(\vec{k}) H(\vec{k})\mathfrak{f}_{\alpha}(\vec{k}) ,\\\nonumber
H(\vec{k})&=
    -2t \begin{pmatrix}
    0&\cos \vec{k}\cdot\vec{R}_1/2 & \cos\vec{k}\cdot\vec{R}_2/2\\
    \cos \vec{k}\cdot\vec{R}_1/2 & 0& \cos\vec{k}\cdot\vec{R}_3/2\\
    \cos \vec{k}\cdot\vec{R}_2/2 & \cos\vec{k}\cdot\vec{R}_3/2 & 0
    \end{pmatrix},
\\\nonumber\\\nonumber
\mathfrak{f}_{\alpha}(\vec{k})  &=(f_{1\alpha}(\vec{k}),f_{2\alpha}(\vec{k}),f_{3\alpha}(\vec{k}))^T.
\end{alignat}
This dispersion yields a flat band at $\epsilon=2t$ and two bands that intersect at Dirac points at $\vec{K}$ and $\vec{K}'$ with $\epsilon(\vec{K})=\epsilon(\vec{K}')=-t$.
We focus on the low energy excitations near the Dirac
nodes, onto the two bands that
touch at $-t$.
Near the low energy Dirac nodes, we will choose the following basis for eigenvectors at $H(\pm\vec{K})$ (suppressing the $SU(3)$ index),
\begin{alignat}{1}
    (e_1^{\pm})^T&=\pm\frac{1}{\sqrt{3}}(e^{i\pi /3},e^{-i\pi/3},1)\\
    (e_2^{\pm})^T&=(e_1^+)^*,
\end{alignat}
where we have used the lower subscript to denote the band index and the upper one to denote the nodal point.
To write an effective Hamiltonian for these, we restore the $SU(3)$ spin index $\alpha$ and define the continuum fermion fields
\begin{equation}
    \psi_{\alpha,\pm}(\vec{q})=
   \begin{pmatrix}
        \left(e^{\pm}_1\right)_if_{\pm\vec{K}+\vec{q},i,\alpha}\\
        \left(e_2^{\pm}\right)_if_{\pm\vec{K}+\vec{q},i,\alpha}
    \end{pmatrix}
\end{equation}

In this basis, we find the continuum effective Dirac Hamiltonian from the second quantized $H(\pm\vec{K}+\vec{q})-H(\pm\vec{K})$,
\begin{equation}
\mathcal{H}_{Dirac}=v_F\int \frac{d^2\vec{q}}{(2\pi)^2}
\psi_{\alpha a}^{\dagger}(q_x\tau^1+q_y\tau^2)\psi_{\alpha a},
\end{equation}
where $v_F=\sqrt{3}t$.
The first index labels the $SU(3)$ spin, the second index labels the Dirac node, and we have made the spinor index implicit.
\section{Monopole wave functions}\label{app:monopole}
As outlined in the main text, the total monopole wave function in the $SU(6)_f$ can be written in the product basis of the spin $SU(3)_s$ and valley $SU(2)_v$ space.
Grouping the $\phi$ into irreducible representations of $SU(3)_s$ and $SU(2)_v$ will greatly aid in intuition and calculation.
After proper antisymmetrization $\wedge^{3}\mathbb{C}^{6}=\mathbb{C}^{20}$, this is in total a $20$-dimensional basis.
Recall our notation in the main text, that we label each state with a fully antisymmetrized multi-index $A=[A_1,A_2,A_3]$, where $A_i\in\{1,\dots,6\}$.
More explicitly, we have
\begin{alignat}{1}
    \phi_A^{\dagger}&=\mathcal{F}_{A}^{\dagger}\mathcal{M}^{\dagger},\\\nonumber
    \mathcal{F}_A^{\dagger}&=f_{[A_1}^{\dagger}f_{A_2}^{\dagger}f_{A_3]}^{\dagger}.
\end{alignat}
As in the main text, we define
the totally antisymmetric tensor $E_{AB}=E_{[A_1,\dots,A_3][B_1,\dots,B_3]}=\frac{1}{6}\epsilon_{A_1\dots A_3B_1\dots B_3}$, which is an invariant bilinear of the antisymmetric $SU(6)$ representation whose Young tableau has one column and $3$ rows.
The monopole, which transforms in the antisymmetric $SU(6)$ self-conjugate representation, decomposes in $SU(3)_s\times SU(2)_v$ as
\begin{alignat}{1}
    \mathbf{20}_6&=(\mathbf{1}_{3}\otimes \mathbf{4}_{2})\oplus (\mathbf{8}_3\otimes\mathbf{2}_2),\\
    \ytableausetup{smalltableaux,aligntableaux=center}
\ydiagram{1,1,1}&=\left(\;\ydiagram{1,1,1},\ydiagram{3}\;\right)\oplus\left(\;\ydiagram{2,1},\ydiagram{2,1}\;\right).
\end{alignat}
The first manifold of states, $(\mathbf{1}_{3}\otimes \mathbf{4}_{2})$, is simply the product of the $SU(3)_s$ singlet state with the fully symmetric $SU(2)_v$ quartet.
By construction, this state will be an antisymmetric $SU(6)$ representation, as desired.
The second manifold of states $(\mathbf{8}_3\otimes\mathbf{2}_2)$, consisting of two octets, is formed by the product of two \textit{mixed symmetry} states, which have only definite permutation symmetry upon the interchange of two indices. For example, one may have the state $\frac{1}{\sqrt{2}}(\ket{+-}-\ket{-+})\ket{+}\in\mathbf{2}_2$, where $\pm$ labels the $SU(2)_v$ degree of freedom.

Let us label $\sigma_{ij}$ as a mixed symmetry spin state that is antisymmetric under permutation of particles $i$ and $j$ and $\rho_{ij}$ as a mixed symmetry valley state, similarly defined.
Then, there are two possible $\rho_{ij}$ for each $i,j$, where $\frac{1}{\sqrt{2}}(\ket{+-}-\ket{-+})\ket{+}$ would be one of the two $\rho_{12}$.
As is known from particle physics, for fixed $i,j$, each $\sigma_{ij}$ contains $8$ states, coming from the direct product of three $SU(3)$ fundamental representations,
\begin{equation}
    \mathbf{3}\otimes \mathbf{3}\otimes \mathbf{3}=\mathbf{10}\oplus \mathbf{8}\oplus \mathbf{8}\oplus \mathbf{1}.
\end{equation}
The mixed symmetry states are exactly the two baryon octets in particle physics, arising from the combination of three light quark flavors.
The reason there are only two octets in the decomposition is because $\sigma_{13}=\sigma_{12}+\sigma_{23}$ is not independent.
After some calculation, one can find the composition that creates the totally antisymmetric $SU(6)$ octet is given by \cite{griffiths2008introduction}
\begin{equation}
\rho_{12}\otimes(\sigma_{31}+\sigma_{32})+\rho_{23}\otimes(\sigma_{12}+\sigma_{13})+\rho_{31}\otimes(\sigma_{23}+\sigma_{21}).
\end{equation}
In the above we have used 
\begin{equation}
    \begin{pmatrix}
    \sigma_{ij}\\\rho_{ij}
\end{pmatrix}=-\begin{pmatrix}
    \sigma_{ji}\\\rho_{ji}
\end{pmatrix}.
\end{equation}
We can form a basis of monopoles that respects the decomposition $SU(6)_f\rightarrow SU(3)_s\times SU(2)_v)$, and all twenty monopoles (relabeled $\Phi$) are shown in Table~\ref{table:monopole}.

\begin{table}[]
\begin{tabular}{|l|l|}
\hline
Representation                         & Monopole Operator                                                                           \\ \hline
$\mathbf{1}_{3}\otimes \mathbf{4}_{2}$ & $\Phi_{1}^{\dagger}=\phi_{[123]}^{\dagger}$                                                 \\ 
                                       & $\Phi^{\dagger}_2=\phi^{\dagger}_{[456]}$                                                   \\
                                       & $\Phi_{3}^{\dagger}=\phi_{[156]}^{\dagger}-\phi_{[246]}^{\dagger}+\phi_{[345]}^{\dagger}$   \\
                                       & $\Phi_{4}^{\dagger}=\phi_{[234]}^{\dagger}-\phi_{[135]}^{\dagger}+\phi_{[126]}^{\dagger}$   \\ \hline
$\mathbf{8}_3\otimes\mathbf{2}_2$      & $\Phi_{5}^{\dagger}=\phi_{[135]}^{\dagger}+\phi_{[234]}^{\dagger}$                          \\ 
                                       & $\Phi_{6}^{\dagger}=2\phi_{[126]}^{\dagger}+\phi_{[135]}^{\dagger}-\phi_{[234]}^{\dagger}$  \\
                                       & $\Phi_{7}^{\dagger}=\phi_{[125]}^{\dagger}$                                                 \\
                                       & $\Phi_{8}^{\dagger}=\phi_{[235]}^{\dagger}$                                                 \\
                                       & $\Phi_{9}^{\dagger}=\phi_{[236]}^{\dagger}$                                                 \\
                                       & $\Phi_{10}^{\dagger}=\phi_{[136]}^{\dagger}$                                                \\
                                       & $\Phi_{11}^{\dagger}=\phi_{[134]}^{\dagger}$                                                \\
                                       & $\Phi_{12}^{\dagger}=\phi_{[124]}^{\dagger}$                                                \\
                                       & $\Phi_{13}^{\dagger}=-\phi_{[246]}^{\dagger}-\phi_{[156]}^{\dagger}$                        \\
                                       & $\Phi_{14}^{\dagger}=-2\phi_{[345]}^{\dagger}-\phi_{[246]}^{\dagger}+\phi_{[156]}^{\dagger}$ \\
                                       & $\Phi_{15}^{\dagger}=-\phi_{[245]}^{\dagger}$                                               \\
                                       & $\Phi_{16}^{\dagger}=-\phi_{[256]}^{\dagger}$                                               \\
                                       & $\Phi_{17}^{\dagger}=-\phi_{[356]}^{\dagger}$                                               \\
                                       & $\Phi_{18}^{\dagger}=-\phi_{[346]}^{\dagger}$                                               \\
                                       & $\Phi_{19}^{\dagger}=-\phi_{[146]}^{\dagger}$                                               \\
                                       & $\Phi_{20}^{\dagger}=-\phi_{[145]}^{\dagger}$      \\\hline                                        
\end{tabular}
\captionsetup{justification=raggedright}
\caption{\label{table:monopole} The twenty monopole operators in the $SU(3)_s\times SU(2)_v$ basis. Note that $\Phi_{5,\dots,12}$ and $\Phi_{13,\dots,20}$ are related by a $\pi$ rotation in $ SU(2)_{v}$.}
\end{table}

\section{Berry phase and Wannier centers}\label{app:berry}
\subsection{Berry Phase of Discrete Symmetries}

We begin with considering the Berry phase associated with discrete IR symmetries.
Note that the overall phase associated with reflection (and charge conjugation, which is not even a lattice symmetry) is not physically meaningful, as it can be changed by a redefinition of the bare monopole operator $\mathcal{M}\rightarrow e^{i\theta}\mathcal{M}$ \cite{Song_2020}.
While the $U(1)_{top}$ phase associated $\mathcal{CR}$ is meaningful and has been considered in similar models, it is not relevant here as our underlying kagome lattice model has no $\mathcal{C}$ symmetry in the UV at $1/3$ filling.

Now what remains is time reversal symmetry. Similar to \cite{Song_2020}, we introduce a Dirac mass $\overline{\psi}\mu^3\psi$ that corresponds to a valley Hall mass.
Such a mass does not lead to a nontrivial $\mathbb{Z}_2$ topological insulator as our spin symmetry is $SU(3)$ and not $SO(3)$. 
Therefore, the monopoles selected by such valley Hall masses should be invariant under the Kramers time reversal $i\mu^2\mathcal{T}_{IR}$, which is exactly the lattice time reversal $\mathcal{T}$.
This uniquely fixes the $U(1)_{top}$ phase associated with $\mathcal{T}$.
\subsection{Berry Phase of Lattice Symmetries}

In order to decompose the band structure of the mean field ansatz, we will use the techinques outlined in \cite{Song_2019}.
By calculating how rotation acts on the occupied bands at high-symmetry points and comparing it to the spectra of appropriately defined Wannier insulator bands, we can appropriately decompose the spinon bands into a sum (or difference, in the case that fragile topology is present) of Wannier bands.
The Wannier bands can be visualized as positive gauge charges sitting at each Wannier center, which contribute to Berry phases upon translation and rotation.
In general, a charge $q_r$ sitting at a rotation center $r$ leads to the Berry phase under $C_n$ rotation centered at $r$ to be
\begin{equation}
    \theta(C_{n,r})=e^{q_r2\pi i/n}.
\end{equation}

As mentioned in the main text, it is mandatory we include a quantum valley Hall mass in order to split the spinon bands into Kramers sub-bands with well-defined spectra.
We first introduce the mass $\overline{\psi}\mu^3\psi$ (identically for all $SU(3)$ spins), which favors the monopole $\Phi^{\dagger}_1$.
In the original unit cell, we will compare the spinon band representation of the Wannier insulator  centered on the upward triangle ($\Gamma^{\vartriangle}$), centered on the downward triangle ($\Gamma^{\triangledown}$), centered on site ($\Gamma^{\circ}$), and centered on the hexagon ($\Gamma^{\varhexagon}$).
The relevant symmetry operations are the threefold rotations around the upward triangle ($C_3^{\vartriangle}$), downward triangle
($C_3^{\triangledown}$), and hexagon ($C_3^{\varhexagon}$).
Note that $C_6^{\varhexagon}$ is broken down to $C_3^{\varhexagon}$ by our choice of mass. 
The high symmetry points are $\vec{K}$, $\vec{K}'$, and $\vec{\Gamma}$, which are left invariant under $C_3$.
Note that immediately, simply from the $SU(3)$ symmetry, we know the Berry phases are trivial. 
This arises from the fact that the occupied spinon bands $\Gamma^{spinon}$ must decompose as 
\begin{equation}
    \Gamma^{spinon}=3\cdot\left(\Gamma_{\bullet}+\dots\right),
\end{equation}
which is a threefold direct sum of a combination of Wannier bands.
Therefore, the resulting pattern of gauge charges (in addition to a background $-1$ charge on each site, which is insignificant) centered at each upward triangle, downward triangle, and hexagon will be some number $q_{\vartriangle,\triangledown,\varhexagon}\equiv 0\pmod{3}$, leading to no Berry phase as all of the relevant rotation operation are $C_3$.
In particular, this means translations $T_1=(C_3^{\triangledown})^{-1}C^{\vartriangle}_3$ and $T_2=(C_3^{\triangledown})^{-1}(C^{\vartriangle})^2(C_3^{\triangledown})^{-1}$
will also carry no Berry phase.
For completeness, we include the explicit results in Table~\ref{table:bands_berry1}.

\begin{table}[h]
\begin{tabular}{|lllllll|}
\hline
\multicolumn{1}{|l|}{Sym.}              & \multicolumn{1}{l|}{$\Gamma^{\vartriangle}$} & \multicolumn{1}{l|}{$\Gamma^{\triangledown}$} & \multicolumn{1}{l|}{$\Gamma^{\circ}$}           & \multicolumn{1}{l|}{$\Gamma^{\varhexagon}$} & \multicolumn{1}{l|}{$\Gamma^{\vec{K},\vec{K}'}$} & $\Gamma^{\vec{\Gamma}}$ \\ \hline
\multicolumn{1}{|l|}{$C_3^{\vartriangle}$}  & \multicolumn{1}{l|}{$\omega^{2l_1}$}           & \multicolumn{1}{l|}{$p_3\omega^{2l_2}$}         & \multicolumn{1}{l|}{$[1,\omega,\omega^2]$}      & \multicolumn{1}{l|}{$p_1\omega^{2l_3}[1,1]$}  & \multicolumn{1}{l|}{$\oplus^31$}                         & $\oplus^31$                       \\
\multicolumn{1}{|l|}{$C_3^{\triangledown}$} & \multicolumn{1}{l|}{$p_3\omega^{2l_1}$}        & \multicolumn{1}{l|}{$\omega^{2l_2}$}            & \multicolumn{1}{l|}{$p_3[1,\omega,\omega^2]$}   & \multicolumn{1}{l|}{$p_2\omega^{2l_3}[1,1]$}  & \multicolumn{1}{l|}{$\oplus^3e^{\mp 2 \pi i/3}$}         & $\oplus^31$                       \\
\multicolumn{1}{|l|}{$C_3^{\varhexagon}$}   & \multicolumn{1}{l|}{$p_1^*\omega^{2l_1}$}      & \multicolumn{1}{l|}{$p_2^*\omega^{2l_2}$}       & \multicolumn{1}{l|}{$p_1^*[1,\omega,\omega^2]$} & \multicolumn{1}{l|}{$\omega^{2l_3}[1,1]$}     & \multicolumn{1}{l|}{$\oplus^3e^{\pm 2 \pi i/3}$}         & $\oplus^31$                       \\ \hline
\multicolumn{7}{|l|}{$\Gamma^{spinon}=3\Gamma_{l=0}^{\vartriangle}$}                                                                                                                                                                                                                                                     \\ \hline
\end{tabular}
\captionsetup{justification=raggedright}
\caption{\label{table:bands_berry1}. Lattice symmetry representations on the Dirac fermions with a valley mass. We have labeled $p_i=e^{i\vec{q}\cdot\vec{R}_i}$ to be the Bloch phase factor associated with translations at the high symmetry momenta $\vec{q}=\vec{K}$, $\vec{K}'$, or $\vec{\Gamma}$.
We have defined $\omega=e^{2\pi i/3}$ and $l_i$ to be the orbital angular momentum of the Wannier functions.}
\end{table}

It still remains to fix is the Berry phase associated with $C_6$, which must be $\pm 1$ from our above analysis.
To isolate the $C_6$ action, we could introduce a mass $-\overline{\psi}\mu^1\psi$, which favors the monopole $\Phi^{\dagger}_1+\Phi^{\dagger}_2+\Phi^{\dagger}_3+\Phi^{\dagger}_4$.
Introducing $\overline{\psi}\mu^1\psi$ breaks down $T_{1,2}$ into $(T_1+T_2,2T_2-T_1)$, tripling the spatial unit cell while preserving the original hexagon-centered $C_6$. We must focus on the rotation action centered on the hexagonal sites the corner of the unit cell, as those are the only symmetries of our spinon bands.
However, repeating our previous analysis at the high symmetry point $\Gamma$ does not uniquely constrain the $C_6$ Berry phase because multiple Wannier patterns could give rise to the same symmetry representation.

Instead, we can find the Berry phase numerically. In order to do this, we place the spin liquid on a finite torus with $2\pi$ flux uniformly spread. We can then find that the monopoles with definite $C_6$ angular momentum, $\Phi_1^{\dagger}\pm\Phi_2^{\dagger}$, transform under $C_6$ with eigenvalues $\pm 1$.
Note that using a remaining phase ambiguity of the monopoles, we have fixed the relative phase between $\Phi_1^{\dagger}$ and $\Phi_2^{\dagger}$ such that $C_6(\Phi_1^{\dagger})=\Phi_2^{\dagger}$, and both phase choices do not affect the existence of the single nontrivial monopole.

\section{Comments on Monopole-Antimonopole Pairs}
\label{app:pairs}
In this appendix, we analyze the relevance of other operators in the zero flux sector, generated by single monopole-antimonopole pairs $\Phi^{\dagger}\Phi$, to argue that the DSL is a quantum critical point.
In order for the DSL to be an un-fine tuned quantum critical point, these operators must all be irrelevant if they are invariant under all microscopic symmetries.
To begin, the all single monopole-antimonopole pairs are charge neutral under $U(1)_{top}$, so we will label the operators by their global $SU(6)$ representation.
As $\Phi$ and $\Phi^{\dagger}$ transform in the $\mathbf{20}$ representation of $SU(6)$, $\Phi^{\dagger}\Phi$ transforms as
\begin{equation}
\mathbf{20}\otimes\mathbf{20}=\mathbf{1}\oplus\mathbf{35}
\oplus\mathbf{175}\oplus\mathbf{189}
\end{equation}
The singlet $\mathbf{1}$ corresponds to a four fermion term.
The most relevant operator in this channel is chiral and not symmetry allowed, while the next most relevant operator in this channel is symmetryt allowed but likely irrelevant, with scaling dimension $\sim4.35$ from large $N_f$ \cite{Chester2016anomalous}.
The $\mathbf{35}$ transforms like the adjoint and is therefore not symmetry allowed.
Lastly, there are allowed terms in the $\mathbf{175}$ and $\mathbf{189}$ channels, but both of these operators are likely irrelevant, with scaling dimensions approximately $4.41$ and $3.84$, respectively from large $N_f$ \cite{Chester2016anomalous}. Note that there are operators in the $\mathbf{189}$ corresponding to four-fermion operators that have scaling dimension $\sim2.92$ from large $N_f$. In general, these may be UV allowed, so we must assume that they are irrelevant in order for the DSL to be a true critical point. The assumptions we make here are similar to the ones made in in order for there to be a stable $N_f=4$ DSL on the triangular and kagome lattices \cite{he_scipost_2022}, and a DSL critical point on the square lattice \cite{zhang2024diracspinliquidunnecessary}.
\section{The Kagome Lattice Chiral Spin Liquid}
\label{app:chiral}
In this appendix, we consider the condensation of a chiral (quantum Hall) mass $\overline{\psi}{\psi}$, which breaks time reversal and reflection symmetry, generating an intermediate chiral spin liquid (CSL) phase at intermediate $U$.
We remark that while a CSL is found on the kagome lattice for the Heisenberg coupling with chiral ring-exchange terms \cite{wu_prb_2016,xu_prb-2023}, one could imagine such a chiral state emerging through the spontaneous breaking of time reversal and reflection.
The CSL hosts topological order, anyon excitations, and gapless edge states.
Moreover, the induced Chern-Simons term from the singlet mass suppresses monopole proliferation as the monopoles will become electrically charged and linearly confined \cite{pisarski_prd_1986,affleck_nucphysb_1989}.
After integrating out the gapped fermions, the specific topological order generated in this case is described by the $K$-matrix
\begin{equation}
    K_{CSL}=\begin{pmatrix}
        2&1\\1&2
    \end{pmatrix},
\end{equation}
identical to that of the Halperin $(221)$ quantum Hall state \cite{wu_prb_2016}, with three superselection sectors spanned by an anyon $a$ such that $a^3=1$ and $\theta_a=2\pi/3$.
The edge theory contains two copropagating bosonic modes and is captured by the $SU(3)_1$ chiral Wess-Zumino-Witten CFT.

\section{$PSU(3)_s\times PSU(2)_v\times U(1)_{top}$ Anomaly}
\label{app:anom_midIR}
It is also insightful to consider the anomaly when we break $SU(6)\rightarrow SU(3)_s\times SU(2)_v$.
Such a branching pattern is applicable as the action of the UV spin and lattice/valley symmetries often decouple exactly this way when embedded into the IR, including for our kagome lattice ansatz and many other parton models.
Therefore, this can be considered as the effective global symmetry in a mid-IR regime.
We will proceed exactly as in the main text, coupling the QED$_3$ to $SU(3)_s$ gauge field $\mathcal{A}^s$, $SU(2)_v$ gauge field $\mathcal{A}^v$, and a $U(1)_{top}$ gauge field $\mathcal{A}^{top}$. As before, the dynamical $U(1)$ gauge field coupled to the Dirac fermions is denoted by $a$.
Defining the classes $w_2^s\in H^2(M_4,\mathbb{Z}_3)$, $w_2^v\in H^2(M_4,\mathbb{Z}_2)$, and $w_2^{TM}\in H^2(M_4,\mathbb{Z}_2)$ as the Brauer and Stiefel-Whitney classes of the $SU(3)_s$, $SU(2)_v$, and tangent bundles respectively, we have the
cocyle compatibility condition
\begin{equation}
\label{cocycle_general_2}
    \oint\frac{da}{2\pi}+\frac{1}{3}w_2^s+\frac{1}{2}w_2^v+\frac{1}{2}w_2^{TM}\in\mathbb{Z}.
\end{equation}
The cocycle condition restricts fields with charge $1$ under $U(1)$ to be fermionic and transform in the fundamentals of the corresponding spin and valley symmetry groups.
The presence of $w_2^v$ also leads to a fractional quantization of the flux
\begin{equation}
    \oint \frac{1}{2}w_2^v+\frac{d\mathcal{A}^{top}}{2\pi}\in\mathbb{Z}.
\end{equation}
To calculate the 't Hooft anomaly,
one can go through the full calculation as in the pure QED$_3$ case or take Eq.~\eqref{eq:final_anomaly_psu6_p1} and use the symmetry breaking pattern to decompose $p_1[\mathcal{A}^6]=3p_1[\mathcal{A}^v]+2p_1[\mathcal{A}^v]$. 
In both cases, we obtain 
\begin{widetext}
\begin{align}
\label{eq:final_anomaly_psu32}
    \frac{S_{bulk}}{2\pi i}&=-\frac{1}{2}p_1[\mathcal{A}^v]+
    \int\dfrac{d\mathcal{A}^{top}}{2\pi}\wedge\left(\frac{1}{2}\dfrac{d\mathcal{A}^{top}}{2\pi}+\frac{w_2^s}{3}\right)\pmod{1}.
\end{align}
\end{widetext}

Notably, in Eq.~\eqref{eq:final_anomaly_psu32} there is the absence of a
term $\sim w_2^s\cup w_2^v$, which would indicate a mixed anomaly between valley and spin flavor symmetries. This is is a generalization of the parity anomaly of Dirac fermions, and while it would be present in cases $N_f=0\pmod{4}$, we see it is absent here. 
Instead, there is only an anomaly for the $SU(2)_v$ symmetry.
Therefore, we expect that without the monopole state, the Dirac spin liquid should still be able to realize a gapped state with full $SU(3)_s$ spin symmetry.
As before one can still see the full bulk action is time reversal invariant, although there is still mixed time reversal anomalies with the flavor and flux conservation symmetries.

The final term in Eq.~\eqref{eq:final_anomaly_psu32} contains the anomalies involving $U(1)_{top}$.
Specifically, using similar arguments as from the main text, one can see that the magnetic particles of $U(1)_{top}$ are still the Dirac fermions, which transform in the $SU(3)_s$ and $SU(2)_v$ fundamentals.

Quite generally for pure QED$_3$, we see that even after decomposing the $SU(N_f)$ flavor symmetry into a tensor product symmetry group as is common in parton constructions and lattice models, the key aspects of the 
QED$_3$ anomaly structure are still present.
\bibliography{main}

     \end{document}